\documentclass[twocolumn]{autart}

\newtheorem{mydef}{Definition}
\newtheorem{mythm}{Theorem}
\newtheorem{myprob}{Problem}

\newtheorem{myexm}{Example}
\newtheorem{remark}{Remark}

\usepackage{multirow}
\usepackage{booktabs}
\usepackage{array}

\usepackage{graphicx}
\graphicspath{{./Pics/}}
\DeclareGraphicsExtensions{.pdf}
\usepackage{subfigure} 
\usepackage{float}
\usepackage{amsmath}
\usepackage{amsfonts}
\usepackage{epsfig}
\usepackage{arydshln}
\usepackage{algpseudocode}
\usepackage{verbatim}
\usepackage{enumerate}
\usepackage{rotating}
\usepackage{color}
\usepackage{caption}
\usepackage{flushend}
\usepackage{mathrsfs}
\usepackage{amssymb}
\usepackage[ruled,vlined,linesnumbered]{algorithm2e}
\usepackage{cite}
\usepackage{soul}
\usepackage{tikz}
\usepackage{hyperref}
\hypersetup{hidelinks} 
\usepackage{makecell}
\usetikzlibrary{arrows.meta}
\usetikzlibrary{patterns}
\usepackage{tcolorbox}
\usepackage{fancyhdr}
\usepackage{makecell}
\usepackage{ulem}
\def\XY#1{{\textcolor{red}{ {\bf XY:} #1}}}

\begin{document}

	\begin{frontmatter}
		
\title{Model Predictive Monitoring of Dynamical Systems \\ for Signal Temporal Logic Specifications}

		\thanks[footnoteinfo]{This work was supported by the National Natural Science Foundation of China (62061136004, 62173226, 61833012). (Corresponding authors: S. Li and X. Yin.)}
		
		\author[SJTU1,SJTU2]{Xinyi Yu}\ead{yuxinyi-12@sjtu.edu.cn},
		\author[SJTU1,SJTU2]{Weijie Dong}\ead{wjd\_dollar@sjtu.edu.cn},  
		\author[SJTU1,SJTU2]{Shaoyuan Li}\ead{syli@sjtu.edu.cn},
		\author[SJTU1,SJTU2]{Xiang Yin}\ead{yinxiang@sjtu.edu.cn} 
		
		\address[SJTU1]{Department of Automation, Shanghai Jiao Tong University, Shanghai 200240, China.} 
		\address[SJTU2]{Key Laboratory of System Control and Information Processing, Ministry of Education of China, Shanghai 200240, China.}

		
		\begin{abstract} 
			Online monitoring aims to evaluate or to predict, at runtime, whether or not the behaviors of a system satisfy some desired specification. It plays a key role in safety-critical cyber-physical systems. 
			In this work, we propose a new   monitoring approach, called \emph{model predictive monitoring}, for specifications described by Signal Temporal Logic (STL) formulae. Specifically, we assume that the observed state traces are generated by an underlying dynamical system whose model is known but the control law is unknown. 
			The main idea is to use the dynamic of the system to predict future states when evaluating the satisfaction of the STL formulae. To this end, effective approaches for the computation of feasible sets of STL formulae are provided. We show that, by explicitly utilizing the model information of the dynamical system, the proposed online monitoring algorithm can falsify or certify of the specification in advance compared with existing algorithms, where no model information is used. We also demonstrate the proposed monitoring algorithm by several real world case studies.
		\end{abstract}
		
	\end{frontmatter}

\section{Introduction}
\subsection{Background}
Cyber-Physical Systems (CPS) are man-made modern engineering systems involving both computational devices and physical dynamics. Safety is one of the major considerations in the designs of many CPS such as intelligent transportation systems, smart manufacturing systems and medical devices.  
For those  safety-critical systems, it is crucial to determine
whether or not the behaviors of the system satisfy some
desired high-level specifications. 
For example, once we detect that the system has violated or will inevitably violate
some desired specifications, additional corrective actions can be
taken to ensure safety.

Specification-based monitoring is one of the major techniques in evaluating behavior correctness of CPS \cite{bartocci2018specification}.  
It is well-known as a light-weight alternative to formal verification and is particularly applicable to block-box systems with unknown inputs for which formal verification cannot be applied.
In this context, it is usually assumed that the desired behavior of the system is described by a specification formula and the state traces (a.k.a.\ signals) generated by the system are observed by a \emph{monitor} that can issue alarms when the specification is violated. In the past years, numerous algorithms have been developed for monitoring specifications described by, e.g., Linear Temporal Logic (LTL) \cite{eisner2003reasoning}, Metric Temporal Logic (MTL) \cite{thati2005monitoring,dokhanchi2014line} and  Signal Temporal Logic (STL) \cite{donze2013efficient,deshmukh2017robust}.
Recent applications of specification-based monitoring techniques include, e.g.,  
robot systems \cite{bonnah2022runtime}, autonomous vehicles \cite{sahin2020autonomous}, fuel control system \cite{jakvsic2018quantitative}, smart cities \cite{ma2021novel}, Internet of Things \cite{zhao2022astl} and intelligent medicines \cite{qin2020clairvoyant}.

Depending on what information can be utilized by the monitor, the monitoring problem can be categorized as \emph{offline}  (e.g., \cite{donze2013efficient, donze2010robust, fainekos2009robustness})  and \emph{online}  (e.g., \cite{dokhanchi2014line, deshmukh2017robust, ho2014online}). In offline monitoring, it is assumed that the complete signal to evaluate has already been generated and the monitor needs to determine either the Boolean satisfaction or the quantitative satisfaction degree of the complete signal. Such offline technique is usually used in the design phase to evaluate the simulated traces of the system prototype.  On the other hand, when the CPS is operating online, the monitor only observes \emph{partial} state traces that have been generated so far. Therefore, \emph{online monitoring} focuses on evaluating signals in real time during the operation of the system in order to, e.g., issue   alarms or trigger corrective actions. 

In the context of qualitative online monitoring, 
monitor may make the following evaluations on the observed partial signals: 
(i) the specification cannot be satisfied, i.e., there is no future possibility to correct the signal;
(ii) the specification has already been satisfied, i.e.,  the future signal does not matter anymore; or 
(iii) inconclusive, i.e., the signal can be either satisfied or not depending on what will happen in the future. 
Furthermore, in the  quantitative setting,  the monitor may also estimate the possible robustness interval based on the observed partial signals. 
In the past years, numerous algorithms have been developed for online monitoring for specifications described by temporal logic formulae.  For example, the basic setting is to consider monitoring the Boolean satisfaction of LTL formulae \cite{bauer2011runtime,abate2019monitor,mascle2020ltl} or MTL formulae \cite{ho2014online}.  
In \cite{dokhanchi2014line, deshmukh2017robust}, algorithms have been developed for quantitatively monitoring the satisfaction of specifications by using robust semantics of STL formulae.

Most of the aforementioned online monitoring techniques are \emph{model-free} in the sense that the satisfaction of the specification is only evaluated based on the observed signal without considering the dynamic of the  system or without predicting future states. 
In some cases, however, the model of the underlying system,  can provide additional information to accelerate the monitoring process. Let us consider a scenario, where for an observed signal, a model-free monitor may provide inconclusive evaluation since the partial signal can be extended to either satisfiable  or unsatisfiable  signals. However, those satisfiable  continuations may not be feasible physically in the dynamical system. In this scenario, by leveraging the model information of the dynamical system and predicting future states, the monitor can better assert that the specification cannot be satisfied before it is actually violated.  We refer such type of online monitoring process to as the \emph{model predictive monitoring}, which uses model information to predict future states so that the specifications can be better evaluated.   

\subsection{Our Contributions}
In this paper, we propose a new model predictive monitoring approach of dynamical systems for STL.
STL formulae are interpreted over continuous time signals and have the advantage of   quantitatively  evaluating the degree of the satisfaction or violation using robust semantics \cite{maler2004monitoring,lindemann2018control,gilpin2020smooth,salamati2021data,hashimoto2022stl2vec,lindemann2019robust}. 
The monitor aims to issue alarms when the specification has  already or will inevitably be violated. However, different from existing approaches, here we explicitly consider the model information of underlying dynamical system. Specifically, we consider a discrete-time nonlinear system. In order to incorporate the model information into the evaluation of STL formulae, we propose the notion of \emph{feasible sets}, which are the regions of states from which STL formulae can potentially be satisfied considering the system dynamic. Effective algorithms in a dynamic programming manner have been developed for computing feasible sets offline. To monitor the specification in real-time, we propose online monitoring algorithms that correctly combine both the online observed partial signals and the offline computed feasible sets. 
We show that the proposed model predictive monitoring algorithm may predict the violation of the specification in advance compared with existing model-free approaches. Hence, it may leave more time for the system to take corrective actions to ensure safety. 

\subsection{Related Works}
\emph{Model-Based Monitoring for State-Discrete Systems: }
There are already many works along the line of leveraging model information for the purpose of online monitoring in the state-discrete domain during the last decade. 
For example, \cite{zhang2012runtime,leucker2012sliding} introduce predictive semantics for monitoring of untimed LTL specifications for systems that are not black boxes. 
In \cite{pinisetty2017predictive}, the authors proposed a predictive runtime verification framework for systems with timing requirements.
Recently, \cite{ferrando2022bridging} extends the single-model predictive monitoring approach to the multi-model case for both  centralized and compositional settings. In \cite{yoon2021predictive}, the authors introduce a Bayesian intent inference framework leveraging the robot's intent information to predict future positions.

\emph{Model-Based Monitoring for State-Continuous Systems: }
However, there are only a few existing works along this line in the state-continuous domain these years.  For example, \cite{waga2021model} and \cite{ghosh2022offline} use   prior knowledge about the over-approximations of  target systems represented by linear hybrid automata and linear dynamical systems, respectively, to tackle the problem of scattered sampling uncertainties during the monitoring process. In   \cite{momtaz2021predicate}, the authors adopt a lightweight mechanism for incorporating bounds on system dynamics to reduce monitoring overhead, and \cite{abbas2022leveraging} exploits the rough knowledge of dynamics for STL specifications. 
For data-driven STL predictive monitoring,  \cite{qin2020clairvoyant} proposes to  use statistical time-series analysis techniques to  predicate future states; \cite{ma2021predictive} uses Bayesian recurrent neural networks learned from  data to predict future states with uncertainties; \cite{lindemann2022conformal} applies conformal prediction to provide probability guarantees. However, these approaches either use some rough information of the model or consider a purely unknown system dynamic.

\emph{Model Checking for STL Specifications: } 
Our pre-computation of feasible sets is closely related to the model checking problem for STL specifications \cite{roehm2016stl, bae2019bounded, lee2021efficient, yu2022stlmc}.  Specifically, in the model checking problem, one is given a system model with an initial state and the objective is to determine whether or not all possible traces generated by the system model from the initial state satisfy the STL formula. Strategies for model checking usually involve computations of reachable sets, which are related to our approach. However, our problem essentially requires to compute the satisfaction region of an STL formula taking a given prefix trace into account. This particular requirement cannot be handled by existing model checking strategies due to the lack of automata representation of STL formulae. 
Finally, we note that the satisfication of an STL formula can also be characterized by control barrier functions (CBFs); see,  e.g., \cite{lindemann2018control, buyukkocak2022control}. However, CBF only provides a sufficient condition for the  satisfication of the task, and it may be overly conservative for the purpose of online monitoring.



The preliminary  version of some  results in this paper is presented in \cite{yu2022online}. 
Compared with \cite{yu2022online}, the present work has the following  differences. The major difference is that 
\cite{yu2022online} only considers a restrictive fragment of STL formulae in which no overlap of time horizons is allowed for temporal operators. In contrast, this work  handles a more  general case that supports arbitrarily overlapped temporal operators. Due to this more general setting, our approaches for both the online monitoring algorithm  and the offline pre-computation  of feasible sets are also very different.  Furthermore, our generalized approach also supports the  the monitoring problem for the satisfaction of the STL task for all control input sequences, while \cite{yu2022online} can only handle the existential case.  Finally, the present work provides four detailed case studies of real-world systems to illustrate the proposed algorithm, which are not provided in  \cite{yu2022online}.

\subsection{Organization} 
The rest of the paper is organized as follows. 
We present some basic preliminaries in Section~\ref{sec:pre} and formulate the problem in Section~\ref{sec:prob}.  Section~\ref{sec:online} presents the main body of the  online monitoring algorithm, which uses feasible sets that are computed offline in  Section~\ref{sec:offline}.  
Section~\ref{sec:extension} is the extension of monitoring satisfaction of STL formula.
The overall framework is demonstrated by  several real world case studies in Section~\ref{sec:case} and finally, we conclude this work in Section~\ref{sec:con}.

\section{Preliminary}\label{sec:pre}
\subsection{System Model}

We consider a discrete-time control system of form
\begin{equation}\label{eq:system}
	x_{k+1} = f(x_k, u_k),
\end{equation}
where 
$x_k \!\in\! \mathcal{X} \ \subset \ \mathbb{R}^n$ is the state at time $k$, 
$u_k \!\in\! \mathcal{U} \ \subset \ \mathbb{R}^m$ is the control input at time $k$
and 
$f:\mathcal{X}\times \mathcal{U} \to \mathcal{X}$ is the dynamic function of the system, which is assumed to be continuous in $\mathcal{X} \times \mathcal{U}$. 
Throughout the paper, we assume that the state space  $\mathcal{X}$ and input space $\mathcal{U}$ are both bounded due to physical constraints.

Suppose that the system is in state $x_k \!\in\!  \mathcal{X}$ at time instant $k\!\in\!\mathbb{Z}_{\geq 0}$.  Then given a sequence of control inputs
$\mathbf{u}_{k:T-1}=  u_k u_{k+1} \dots u_{T-1}  \!\in\! \mathcal{U}^{T-k}$,
the  solution of the system is a sequence of states
$\xi_f(x_k,\mathbf{u}_{k:T-1}) = {\mathbf{x}_{k+1:T} = } x_{k+1} \dots x_{T}\!\in\! \mathcal{X}^{T-k}$ such that $x_{i+1}=f(x_i,u_i), i=k,\dots, T-1$.

\subsection{Signal Temporal Logic}
We use Signal Temporal Logic (STL) formulae  with bounded-time temporal operators \cite{maler2004monitoring} to describe whether or not the trajectory of the system satisfies some desired high-level properties. 
Formally, the syntax of STL formulae is as follows 
\[
\Phi ::=   \top\mid \pi^\mu \mid \neg \Phi \mid \Phi_1 \wedge \Phi_2 \mid \Phi_1 \mathbf{U}_{[a,b]} \Phi_2,
\]
where $\top$ is the \textsf{true} predicate, 
$\pi^\mu$ is an atomic predicate whose truth value is determined by the sign of its underlying predicate function $\mu:\mathbb{R}^n \to \mathbb{R}$  and it is true at state $x_k$ when $\mu(x_k) \geq 0$; otherwise it is false.
Notations $\neg$ and $\wedge$ are the standard Boolean operators ``negation" and ``conjunction", respectively,  which can further induce ``disjunction" by $\Phi_1 \vee \Phi_2:=\neg(\neg \Phi_1 \wedge \neg \Phi_2)$
and ``implication" by $\Phi_1 \to \Phi_2:= \neg \Phi_1 
\vee   \Phi_2$. 
$\mathbf{U}_{[a,b]}$ is the temporal operator ``\emph{until}", where $a,b\in \mathbb{Z}_{\geq 0}$ are two integer instants with $a \leq b$.
As we consider a discrete-time setting, we use $[a,b]$ as a shorthand notation for the discrete-time interval $[a,b]\cap \mathbb{Z}$ which is the set of all integers between $a$ and $b$ including $a$ and $b$; this set is non-empty when $a,b\in \mathbb{Z}_{\geq 0}$ and $a \leq b$.
Also we note that, since we consider a discrete-time setting, time instant $k \in \mathbb{Z}_{\geq 0}$ does not necessarily represent the real time; the specific real time also depends on the sampling rate when discretizing the system.

STL formulae are evaluated on state sequence $\mathbf{x}=x_0x_1\cdots$.  
We use notation $(\mathbf{x},k) \models \Phi$ to denote that sequence $\mathbf{x}$ satisfies STL formula $\Phi$ at time instant $k$. 
The reader is referred to \cite{maler2004monitoring} for more details on the semantics of STL formulae. 
Particularly, we have 
$(\mathbf{x},k)\models \pi^\mu$ iff $\mu(x_k)\geq 0$, i.e., $\mu(x_k)$ is non-negative for the current state $x_k$,  
and 
$(\mathbf{x},k)\models \Phi_1 \mathbf{U}_{[a,b]} \Phi_2$ 
iff $\exists k' \!\in\! [k+a, k+b]$ such that $(\mathbf{x},k') \models \Phi_2$
and  $\forall k'' \!\in\! [k, k']$, we have $(\mathbf{x},k'') \models \Phi_1$, i.e., 
$\Phi_2$ will hold at some instant between $[k+a, k+b]$ in the future and before that $\Phi_1$ always holds. 
Furthermore, 
we can also induce temporal operators\vspace{-6pt}
\begin{itemize}
	\item 
	``\emph{eventually}" $\mathbf{F}_{[a,b]} \Phi:= \top \mathbf{U}_{[a,b]} \Phi$
	such that it holds when $(\mathbf{x},k) \models \Phi$ for some $k'\!\in\! [k+a,k+b]$; and \medskip
	\item 
	``\emph{always}" $\mathbf{G}_{[a,b]} \Phi:=\neg \mathbf{F}_{[a,b]} \neg \Phi$ such that it holds  when $(\mathbf{x},k) \models \Phi$ for any $k'\!\in\! [k+a,k+b]$. \vspace{-6pt}
\end{itemize}
We write $\mathbf{x} \models \Phi$ whenever $(\mathbf{x},0) \models \Phi$.

Given an STL formula $\Phi$, in fact, it is well-known that the satisfaction of  $\Phi$ can be completely determined only by those states within its \emph{horizon}.   
Specifically, we will use notation $\Phi^{[S,T]}$ to emphasize that the satisfaction of formula $\Phi$ only depends on time horizon  $[S,T]$, 
where   $S$ is the starting instant of $\Phi$ which is the minimum time instant that appears in the formula 
and $T$ is the terminal instant of $\Phi$ which is the maximum sum of all nested upper bounds.
For example, for $\Phi = \mathbf{F}_{[2,7]}\pi^{\mu_1} \wedge  \mathbf{G}_{[3,12]} \pi^{\mu_2}$, we have $T = \max\{7,12\} = 12$ and $S = \min\{2, 3\} = 2$.

\section{Problem Formulation}\label{sec:prob}
\subsection{Fragment of STL Formulae}
In this paper, we consider the following restricted but still expressive enough fragments of STL formulae:
\begin{subequations} \label{eq:stl}
	\begin{align}
		\varphi ::= \top \mid \pi^\mu \mid \neg \varphi \mid \varphi_1 \wedge \varphi_2, \qquad \qquad  \\
		\Phi::= \mathbf{F}_{[a,b]} \varphi \mid \mathbf{G}_{[a,b]} \varphi \mid \varphi_1\mathbf{U}_{[a,b]} \varphi_2 \mid \Phi_1 \wedge \Phi_2,
	\end{align}
\end{subequations}
where $\varphi_1, \varphi_2$ are formulae of class $\varphi$, and $\Phi_1, \Phi_2$ are formulae of class $\Phi$.  
Specifically, we only allow the temporal operators to be applied once for Boolean formulae. 

\begin{figure}[t]
	\centering
\usetikzlibrary{intersections}
\usetikzlibrary{patterns}
\usetikzlibrary{quotes,angles}
\usetikzlibrary{calc}
\usetikzlibrary{decorations.pathreplacing}

\def\unit{0.5}

\begin{tikzpicture}
	\foreach \x in {0,1,2,...,15,16}
	\draw (\x*\unit, 2.5-0.03) -- (\x*\unit, 2.5+0.03);
	\foreach \x in {0,1,2,...,15}
	\draw (\x*\unit+0.5*\unit, 2.5 + 0.2) node {\tiny \x};
	\draw[very thick] (-0.3+0.5*\unit, 2.5) -- (7.9+0.5*\unit,2.5);
	\draw[thick] (7.9+0.5*\unit, 2.5) -- (7.75+0.5*\unit, 2.5+0.1);
	\draw[thick] (7.9+0.5*\unit, 2.5) -- (7.75+0.5*\unit, 2.5-0.1);

	\draw (7*\unit+0.5*\unit, 2+0.15) node {\tiny{$\mathbf{G}_{[3,11]} x \!\in\! \mathcal{H}_{\mathbf{G}}$}};
	\draw[thick] (2.5*\unit+0.5*\unit, 2-0.1) -- (2.5*\unit+0.5*\unit, 2+0.1);
	\draw[thick] (11.5*\unit+0.5*\unit, 2-0.1) -- (11.5*\unit+0.5*\unit, 2+0.1);
	\draw[thick] (2.5*\unit+0.5*\unit,2) -- (11.5*\unit+0.5*\unit,2);

	\draw (10*\unit+0.5*\unit, 1.5+0.15) node {\tiny{$\mathbf{F}_{[5,15]} x \!\in\! \mathcal{H}_{\mathbf{F}}$}};
	\draw[thick] (4.5*\unit+0.5*\unit, 1.5-0.1) -- (4.5*\unit+0.5*\unit, 1.5+0.1);
	\draw[thick] (15.5*\unit+0.5*\unit, 1.5-0.1) -- (15.5*\unit+0.5*\unit, 1.5+0.1);
	\draw[thick] (4.5*\unit+0.5*\unit,1.5) -- (15.5*\unit+0.5*\unit,1.5);

	\draw (11*\unit+0.5*\unit, 1+0.15) node {\tiny{$x \!\in\! \mathcal{H}^1_{\mathbf{U}} \mathbf{U}_{[8,14]} x \!\in\! \mathcal{H}^2_{\mathbf{U}}$}};
	\draw[thick] (7.5*\unit+0.5*\unit, 1-0.1) -- (7.5*\unit+0.5*\unit, 1+0.1);
	\draw[thick] (14.5*\unit+0.5*\unit, 1-0.1) -- (14.5*\unit+0.5*\unit, 1+0.1);
	\draw[thick] (7.5*\unit+0.5*\unit,1) -- (14.5*\unit+0.5*\unit,1);

	\draw (4, 0.5) node {\small{(a)}};

	\foreach \x in {0, 1,2,...,15,16}
	\draw (\x*\unit, -0.2-0.03) -- (\x*\unit, -0.2+0.03);
	\foreach \x in {0,1,2,...,15}
	\draw (\x*\unit+0.5*\unit, -0.2 + 0.2) node {\tiny \x};
	\draw[very thick] (-0.3+0.5*\unit,-0.2) -- (7.9+0.5*\unit,-0.2);
	\draw[thick] (7.9+0.5*\unit, -0.2) -- (7.75+0.5*\unit, -0.2+0.1);
	\draw[thick] (7.9+0.5*\unit, -0.2) -- (7.75+0.5*\unit, -0.2-0.1);

	\draw (1*\unit+0.5*\unit, -0.7+0.15) node {\tiny{$\mathbf{G}_{[0,2]} x \!\in\! \mathcal{H}^1_{\mathbf{U}}$}};
	\draw (5*\unit+0.5*\unit, -0.7+0.15) node {\tiny{$\mathbf{G}_{[3,7]} x \!\in\! \mathcal{H}^1_{\mathbf{U}} \cap \mathcal{H}_{\mathbf{G}}$}};
	\draw (9.5*\unit+0.5*\unit, -0.7+0.15) node {\tiny{$\mathbf{G}_{[8,11]} x \!\in\! \mathcal{H}_{\mathbf{G}}$}};
	\draw[thick] (-0.5*\unit+0.5*\unit, -0.7-0.1) -- (-0.5*\unit+0.5*\unit, -0.7+0.1);
	\draw[thick] (2.5*\unit+0.5*\unit, -0.7-0.1) -- (2.5*\unit+0.5*\unit, -0.7+0.1);
	\draw[thick] (7.5*\unit+0.5*\unit, -0.7-0.1) -- (7.5*\unit+0.5*\unit, -0.7+0.1);
	\draw[thick] (11.5*\unit+0.5*\unit, -0.7-0.1) -- (11.5*\unit+0.5*\unit, -0.7+0.1);
	\draw[thick] (-0.5*\unit+0.5*\unit,-0.7) -- (11.5*\unit+0.5*\unit,-0.7);

	\draw (10*\unit+0.5*\unit, -1.2+0.15) node {\tiny{$\mathbf{F}_{[5,15]} x \!\in\! \mathcal{H}_{\mathbf{F}}$}};
	\draw[thick] (4.5*\unit+0.5*\unit, -1.2-0.1) -- (4.5*\unit+0.5*\unit, -1.2+0.1);
	\draw[thick] (15.5*\unit+0.5*\unit, -1.2-0.1) -- (15.5*\unit+0.5*\unit, -1.2+0.1);
	\draw[thick] (4.5*\unit+0.5*\unit, -1.2) -- (15.5*\unit+0.5*\unit, -1.2);

	\draw (11*\unit+0.5*\unit, -1.7+0.15) node {\tiny{$x \!\in\! \mathcal{H}^1_{\mathbf{U}} \mathbf{U}^{\prime}_{[8,14]} x \!\in\! \mathcal{H}^2_{\mathbf{U}}$}};
	\draw[thick] (7.5*\unit+0.5*\unit, -1.7-0.1) -- (7.5*\unit+0.5*\unit, -1.7+0.1);
	\draw[thick] (14.5*\unit+0.5*\unit, -1.7-0.1) -- (14.5*\unit+0.5*\unit, -1.7+0.1);
	\draw[thick] (7.5*\unit+0.5*\unit, -1.7) -- (14.5*\unit+0.5*\unit, -1.7);

	\draw (4, -2.2) node {\small{(b)}};

\end{tikzpicture}
	\caption{Illustration of formulae equivalence in  Example~1.}
	\label{fig:equiv}
\end{figure}
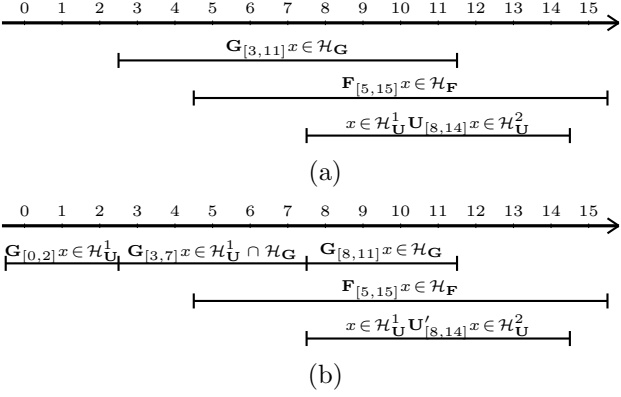

Note that, for the standard ``until" operator, 
$\varphi_1\mathbf{U}_{[a,b]} \varphi_2 $ requires that 
$\varphi_1$ holds \emph{from the initial instant} before $\varphi_2$ holds. 
In order to facilitate subsequent expression, we introduce a new temporal operator 
$\mathbf{U}'$ defined by  $(\mathbf{x},k)\models \Phi_1 \mathbf{U}'_{[a,b]} \Phi_2$ 
iff $\exists k' \!\in\! [k+a, k+b]$ such that $(\mathbf{x},k') \models \Phi_2$
and  $\forall k'' \!\in\! [k+a, k']$, we have $(\mathbf{x},k'') \models \Phi_1$.
Compared with $\mathbf{U}$, the new operator $\mathbf{U}'$ only required that $\Phi_1$ holds \emph{from  instant $a$} before $\Phi_2$ holds. 
Throughout this paper, we will refer ``$\mathbf{U}'$" to as the ``until" operator. 
As illustrated by Figure~\ref{fig:equiv}, our setting is without loss of generality since we can express the standard $\mathbf{U}$ using  $\mathbf{U}'$ by:  
\[
(\mathbf{x}, k) \!\models\! \Phi_1 \mathbf{U}_{[a,b]} \Phi_2
\Leftrightarrow	(\mathbf{x}, k) \!\models\! (\Phi_1 \mathbf{U}'_{[a,b]} \Phi_2) \wedge (\mathbf{G}_{[0, a]} \Phi_1).
\] 
Intuitively, the effective horizon of $\mathbf{U}_{[a,b]}$ is $[0,b]$, while the effective horizon of $\mathbf{U}'_{[a,b]}$ is $[a,b]$. This separation is mainly technical  and  for the sake of efficiency. As we will show later, handling ``until" is more complicated than handling ``always", and therefore, it is beneficial to have ``until" with short  effective horizon.


Furthermore, we can always rewrite Boolean formula $\varphi$ in Eq.~(\ref{eq:stl}a) in terms of the region of states satisfying the formula.
Specifically, for  predicate $\pi^\mu$, its satisfaction region is the solution of inequality $\mu(x) \!\geq\! 0$;  we denote it by set $\mathcal{H}^{\mu}$, i.e., 
$\mathcal{H}^{\mu} = \{x \!\in\! \mathcal{X} \mid \mu(x) \geq 0 \}$. 
Similarly, we have 
$\mathcal{H}^{\neg \varphi} = \mathcal{X} \setminus \mathcal{H}^{\varphi}$ and $\mathcal{H}^{\varphi_1 \wedge \varphi_2} = \mathcal{H}^{\varphi_1} \cap \mathcal{H}^{\varphi_2}$.
Hereafter, instead of using   $\varphi$, 
we will only write it  as $x \!\in\! \mathcal{H}^\varphi$ or simply  $x \!\in\! \mathcal{H}$  using its satisfaction region. 

Based on the above discussion,  STL formulae $\Phi$ in Eq.~(\ref{eq:stl}) can be expressed equivalently by:
\begin{align}\label{eq:stl'}
&\Phi  ::=  \\
&  \mathbf{F}_{[a,b]} x \!\in\! \mathcal{H} \mid \mathbf{G}_{[a,b]} x \!\in\! \mathcal{H} \mid   
  x \!\in \mathcal{H}^1 \mathbf{U}'_{[a,b]} x \!\in\! \mathcal{H}^2 \mid \Phi_1 \wedge \Phi_2, \nonumber
\end{align}
where $\mathcal{H}\subseteq \mathbb{R}^n$ is a set of states representing the satisfaction region of a Boolean formula. 

In summary, we  consider STL formulae of  form\vspace{-6pt}
\begin{equation}\label{eq:stl-seg}
    \Phi = \bigwedge_{i=1}^{N} \Phi_i^{[a_i, b_i]}, \vspace{-6pt}
\end{equation}
where 
$\Phi_i^{[a_i, b_i]}$  is a sub-formula that applies within time interval $[a_i, b_i]$ in the form of 
$\mathbf{G}_{[a_i, b_i]} x \!\in\! \mathcal{H}_i$, $\mathbf{F}_{[a_i, b_i]} x \!\in\! \mathcal{H}_i$ or $x \!\in\! \mathcal{H}_{i}^1 \mathbf{U}^{\prime}_{[a_i, b_i]} x\! \in\! \mathcal{H}_{i}^2$, and $N$ denotes the total number of sub-formulae. 

We denote by $\mathcal{I}=\{1,\dots,N\}$ the index set of all sub-formulae. 
We assume that the indices    are ordered according to the starting instants of the sub-formulae, i.e., $a_1\leq a_2\leq\cdots\leq a_N$.  
For each sub-formula $i \!\in\! \mathcal{I}$, we denote by $O_i \!\in\! \{\mathbf{G}, \mathbf{F},\mathbf{U}'\}$ the unique temporal operator in $\Phi_i$.  
Note that, for each time instant $k$, there may have multiple sub-formulae applied, 
and we denote by   $\mathcal{I}_k  = \{i \!\in\! \mathcal{I} \mid a_i\leq k \leq b_i  \}$  the index set of sub-formulae that are effective at instant $k$. 
Similarly, we denote by 
$\mathcal{I}_{<k} = \{i \!\in\! \mathcal{I} \mid b_i<k\}$
and 
$\mathcal{I}_{>k} = \{i \!\in\! \mathcal{I} \mid k< a_i\}$
the index sets of sub-formulae that are effective strictly before and after instant $k$, respectively.

\begin{myexm}\upshape
Let us consider the following  STL formula of the original form~\eqref{eq:stl}   with horizon  $T = 15$
\begin{equation}\label{eq:example}
	\Phi\!=  \mathbf{G}_{[3,11]} x \!\in\! \mathcal{H}_\mathbf{G}  \wedge  \mathbf{F}_{[5,15]} x \!\in\! \mathcal{H}_\mathbf{F} 
	\wedge  x \!\in\! \mathcal{H}^1_\mathbf{U} \mathbf{U}_{[8,14]} x \!\in\! \mathcal{H}^2_\mathbf{U}.
\end{equation} 
The effective horizon of each sub-formula is shown in Figure~\ref{fig:equiv}(a).
Equivalently, this formula can also be written  in the form of~\eqref{eq:stl'} as
\begin{align}\label{eq:example-prim}
	&  \Phi = ( \mathbf{G}_{[0,2]} x \!\in\! \mathcal{H}^1_\mathbf{U}) 
	\wedge 
	( \mathbf{G}_{[3,7]} x \!\in\! \mathcal{H}^1_\mathbf{U} \cap \mathcal{H}_\mathbf{G} ) 
	\wedge	
	\\	
	& (\mathbf{F}_{[5,15]} x \!\in\! \mathcal{H}_\mathbf{F})\wedge 
	( \mathbf{G}_{[8,11]} x \!\in\! \mathcal{H}_\mathbf{G} ) 
	\wedge
	(x \!\in\! \mathcal{H}^1_\mathbf{U} \mathbf{U}^{\prime}_{[8,14]} x \!\in\! \mathcal{H}^2_\mathbf{U})\nonumber
\end{align}
which is  shown in Figure~\ref{fig:equiv}(b).
In this case, we have $\mathcal{I} = \{1,2,3,4,5\}$ and $O_1, O_2,O_4 = \mathbf{G}, O_3 = \mathbf{F}$ and $ O_5 = \mathbf{U}'$. 
Also, for time instant $k=9$, we have $\mathcal{I}_k=\{3,4,5\}$.
\end{myexm}

\subsection{Online Monitoring of STL}\label{subsec:onlinemonitoring}
Given a  state sequence $\mathbf{x}$, whose length is equal to or longer than the horizon of $\Phi$, we can always completely determine whether or not $\mathbf{x}\models \Phi$. 
However, during the operation of the system, at each time $k$, we can observe the current state $x_k$, and therefore,  only the partial signal    $\mathbf{x}_{0:k}=x_0x_1\cdots x_k$ (called \emph{prefix}) is available at time instant $k$, and the remaining signals $\mathbf{x}_{k+1: T}$   (called \emph{suffix}) will only be available in the future. 
We say a prefix signal $\mathbf{x}_{0:k}$ is\vspace{-6pt}
\begin{itemize}
    \item 
    \emph{violated} if for any control input $\mathbf{u}_{k: T-1}$, we have $\mathbf{x}_{0:k} \xi_f(x_k, \mathbf{u}_{k:T-1}) \not\models \Phi$; 
    \medskip
	\item 
    \emph{feasible} if for some control input $\mathbf{u}_{k: T-1}$, we have $\mathbf{x}_{0:k} \xi_f(x_k, \mathbf{u}_{k:T-1}) \models \Phi$. 
\end{itemize} 
Intuitively, a prefix signal is violated if we know for sure in advance that the formula will be violated inevitably. 
For example, for safety specification $\mathbf{G}_{[0,T]}x\!\in\! \mathcal{H}$, once the system reaches a state $x_k\!\notin\! \mathcal{H}$ for $k<T$, we know immediately that the formula is violated. 
Also, if the system is in state $x_k$ from which no solution $\xi_f(x_k,\mathbf{u}_{k:T-1})$ can be found such that each state is in region $\mathcal{H}$, then we can also claim the formula cannot be satisfied anymore, i.e., it is violated. 

Therefore, an online  \emph{monitor} is a function 
\[
\mathcal{M}: \mathcal{X}^* \to \{0,1\}
\] 
that determines the satisfaction of formula based on the partial signal, where $\mathcal{X}^*$ denotes the set of all finite sequences over $\mathcal{X}$, ``$0$" denotes ``feasible" and ``$1$" denotes ``violated".   
Then the online monitoring problem is formulated as follows.  

\begin{myprob}\label{prob}\upshape
Given a dynamical system of form \eqref{eq:system} and an STL formula $\Phi$ as in \eqref{eq:stl-seg}, design 
an online monitor $\mathcal{M}: \mathcal{X}^* \to \{0,1\}$ such that 
for any prefix signal $\mathbf{x}_{0:k}$ where $k \leq T$, we have 
$\mathcal{M}(\mathbf{x}_{0:k}) = 1$ iff $\mathbf{x}_{0:k}$ is a violated prefix. 
\end{myprob}

\begin{remark}\label{remark:direct}\upshape
We note that, for any prefix signal $\mathbf{x}_{0:k}$,  it is 
a violated prefix iff we cannot find a sequence of control inputs $\mathbf{u}_{k:T-1}$ such that $\mathbf{x}_{0:k} \xi_{f}(x_k, \mathbf{u}_{k:T-1})  \models \Phi$.
The existence of such a control sequence can be determined by the binary encoding technique proposed in \cite{raman2014model}.  Therefore, a naive approach for designing  an online monitor is to solve the above constrained satisfaction problem based on $\mathbf{x}_{0:k}$. However, such a direct approach has the following issues \vspace{-6pt}
\begin{itemize}
    \item 
    First, the computations are performed purely online by solving a satisfaction problem, which is computationally very challenging especially for nonlinear systems with long horizon STL formulae.  Hence, the monitor may not be able to provide  evaluations in time.      
	\medskip
    \item 
    Second, this requires to  store the entire state sequence up to now. 
    It is more desirable if the monitor can just store the satisfaction status of the formula by ``forgetting" those irrelevant information. \vspace{-6pt}
\end{itemize}
Compared with the direct approach, in this paper, we will present an alternative approach by \emph{pre-computing} the set of feasible regions in an offline fashion. Then the pre-computed information will be used online, which ensures timely online evaluations.  
\end{remark}

\begin{remark}\upshape
Problem~1 is formulated by assuming that the control inputs are unknown. 
There are several motivations for this problem setting. Note that, an online monitor is usually implemented as an \emph{additional component} at the top of a control system, 
which can be either semi-autonomous  in the sense that it is controlled by an experienced human-operator  or a black-box controller in the sense that the control law is claimed to be ``correct" but cannot be formally verified. For both cases,  we do not know precisely the control inputs of the system. For the purpose of safety, one still wants to add a monitor at the top of the system.  When the control law is known, the system dynamic is already autonomous. For this case, however, the unknown control inputs here can be considered as disturbances from environments.  Therefore, our problem formulation can also be applied to the case of known control law under unknown disturbances.
\end{remark}

\section{Set-Based Online Monitoring} \label{sec:online}

\subsection{Remaining Formulae and Feasible Set}
As we mentioned above, our objective is to evaluate the satisfaction of STL formulae of the form~\eqref{eq:stl-seg} which is the conjunction of several sub-formulae. Specifically, at each instant $k$, the monitor needs to determine the following two issues:\vspace{-6pt}
\begin{itemize}
	\item
	for  sub-formulae effective currently,  check whether or not each of them has been achieved;  and \medskip
	\item 
	for those sub-formulae (either effective currently or in the future) that have not  been achieved, 
	check whether or not the system is still able to fulfill them in the future. 
\end{itemize}

To formalize the above  issues, we use $I\subseteq \mathcal{I}$ to denote the index set of the remaining sub-formulae (which will be referred to as the \emph{remaining set} latter), i.e., sub-formulae that have not been achieved yet. Then  we introduce the notion of \emph{$I$-remaining formulae} as follows.

\begin{mydef}[$I$-remaining formula]\label{def:I-formula}\upshape 
Given an STL formula $\Phi$ of form~\eqref{eq:stl-seg}, 
a subset of indices $I\subseteq \mathcal{I}$ and a time instant $k\in [0,T]$,  $I$-remaining formula at instant $k$ is defined by 
\begin{equation}
	\hat{\Phi}_{k}^I = \bigwedge_{i   \in I\cap \mathcal{I}_{k}   } \Phi_i^{[k,b_i]} \wedge \bigwedge_{i \in I\cap\mathcal{I}_{>k}   } \Phi_i^{[a_i,b_i]}, \vspace{-6pt}
\end{equation}	
where $\Phi_i^{[k,b_i]}$ is obtained from $\Phi_i^{[a_i,b_i]}$ by replacing the start instant of the temporal operator from $a_i$ to $k$.
\end{mydef}

Intuitively, $\hat{\Phi}_{k}^I $ denotes the conjunction of all sub-formulae that have not been achieved.  
Clearly,  sub-formulae with index in $\mathcal{I}_{>k}$ effective in the future are naturally  not achieved. 
For sub-formulae with index in $\mathcal{I}_{k}$ effective currently, we only consider those in $I$. 
Furthermore, since we are only interested in the satisfiability in the future,  the formulae is truncated from the given instant $k$.

\begin{myexm}[Cont.]\label{exm:monitor} \upshape
Let us consider the STL formula $\Phi$ in Equation~\eqref{eq:example-prim}. 
For time instant $k=7$, we have $\mathcal{I}_{<7} = \{1\}, \mathcal{I}_7 = \{2,3\}$ and $\mathcal{I}_{>7} = \{4,5\}$. 
Now, suppose that the remaining index set at $k=7$ is $I=\{3,4,5\}$, which is determined by the sequence $\mathbf{x}_{0:6}$. 
Then $I$-remaining formula at $k=7$ is
	$\hat{\Phi}^{I}_7= 
	(\mathbf{F}_{[7,15]} x \!\in\! \mathcal{H}_\mathbf{F})
	\wedge
	( \mathbf{G}_{[8,11]} x \!\in\! \mathcal{H}_\mathbf{G} ) 
	\wedge
	(x \!\in\! \mathcal{H}^1_\mathbf{U} \mathbf{U}^{\prime}_{[8,14]} x \!\in\! \mathcal{H}^2_\mathbf{U})$.
\end{myexm}

In order to capture whether or not the $I$-remaining  formulae can possibly be fulfilled in the future under the constraint of the system dynamic,  we introduce the notion of $I$-remaining feasible set. 

\begin{mydef}[$I$-remaining feasible set]\label{def:feasibleset}\upshape
Given an STL formula $\Phi$ of form~\eqref{eq:stl-seg}, 
a subset of indices $I\subseteq \mathcal{I}$ and a time instant $k\in [0,T]$, 
the $I$-remaining feasible set at instant $k$, denoted by 
$X_k^{I} \subseteq \mathcal{X}$, is the set of states from which there exists a solution that satisfies the $I$-remaining formula at $k$, i.e.,
\begin{align}\label{eq:feasibleset}
	X_k^{I} \! =\! 
	\left\{ 
	x_k \in \mathcal{X} \,\middle\vert\, \!\!\!\!
	\begin{array}{cc}
		\exists \ \mathbf{u}_{k:T-1} \in \mathcal{U}^{T-k} \\
		\text{ s.t. } x_k \xi_f(x_{k}, \mathbf{u}_{k:T-1}) \models \hat{\Phi}_{k}^I
	\end{array} 
	\right\}.\vspace{-6pt}
\end{align}
\end{mydef}

In what follows, we will present the main online monitoring algorithm by using the $I$-remaining feasible sets. 
The computation  of set $X_k^{I}$ will be detailed in Section~\ref{sec:offline}.

\subsection{Online Monitoring Algorithm}\label{subsec:onlinealg}
Note that, although we consider temporal operator ``Eventually''  in the semantics, it is subsumed by operator  ``Until'' since 
$\mathbf{F}_{[a_i,b_i]} x \!\in\! \mathcal{H}_i$ can be expressed as $x \!\in\! \mathcal{X} \mathbf{U}'_{[a_i,b_i]}x \!\in\! \mathcal{H}_i$. 
Therefore, technically, we only need to handle  temporal operators $\mathbf{G}$ and $\mathbf{U}'$. 
Specifically, in terms of the satisfaction: \vspace{-6pt}
\begin{itemize}
	\item 
	For sub-formula of form $\Phi_i^{[a_i,b_i]}=\mathbf{G}_{[a_i,b_i]}x \!\in\! \mathcal{H}_i$ with operator $\mathbf{G}$, 
	it is satisfied only when state $x_k$ is still in the region $\mathcal{H}_i$ at the last instant $b_i$.
	\medskip
	\item 
	For sub-formula of form $\Phi_i^{[a_i,b_i]}=x \!\in\! \mathcal{H}_i^1\mathbf{U}'_{[a_i,b_i]}x \!\in\! \mathcal{H}_i^2$, however, its satisfaction can be determined at any instant  $k  \!\in\! [a_i, b_i]$ 
	if state $x_k$ is in region $\mathcal{H}_i^1 \cap \mathcal{H}_i^2$.
\end{itemize}

 
Based on the above discussion, now we present the  complete online monitoring algorithm, which is shown in  Algorithm~1.  
We use a global variable $I$ to record the indices of sub-formulae that have not been satisfied. 
We start from the initial instant $k=0$ (line~1) and $I$ is set as the indices of all formulae  $\mathcal{I}$ (line~2). 
The monitor decision for each instant $k$ is computed in the while-loop. 
Specifically, the monitor first reads the current state $x_k$ (line~4) and 
uses its $I$-remaining  feasible set $X_k^{I}$  to issue a monitoring decision. 
Specifically, if $x_k$ is not in $X_k^{I}$, we know that entire formula cannot be satisfied anymore  (lines~5-7).
If it is in $X_k^{I}$, then we use this state information to determine  whether or not some sub-formulae in set $I$ are achieved based on  rules discussed above (lines~8-12).  
If sub-formulae $\Phi_i$ is satisfied, then we delete its index $i$ from the remaining set $I$ (line~13).
Note that we only need to check the satisfaction of remaining sub-formulae with indices in  $I\cap \mathcal{I}_k$ since sub-formulae with indices in $\mathcal{I}_{>k}$ cannot be satisfied at instant $k$. 
This process is repeated until remaining index set $I$ is empty, i.e., the entire formula is satisfied. 

\IncMargin{1em}
\begin{algorithm}[ht]
	\caption{Online Monitoring Algorithm}
	\label{alg:online}
	\KwIn{feasible sets}
	\KwOut{monitoring decision $\mathcal{M}_k$}
	$k \gets 0$ \\
	$I \gets \mathcal{I} $ \\
	\While{$I \neq \emptyset$}
	{
		read new current state $x_k$ \\ 
		
		\If {$x_k \!\notin\! X_k^{I} $ }
		{
		    $\mathcal{M}_k = 1$ \\
			\textbf{return} ``\textit{prefix is violated}''
		}
		\Else
		{
		    $\mathcal{M}_k = 0$\\
			\ForAll{$i \!\in\! I\cap \mathcal{I}_k$}
			{
				\If {\emph{[$O_i \!=\! \mathbf{G}  \wedge  k\!=\!b_i  \wedge  x_k \!\in\! \mathcal{H}_i$]  
				\textbf{or}\\ {  }  \quad [$ O_i \!=\! \mathbf{U}'  \wedge  x_k \!\in\! \mathcal{H}_{i}^1 \cap \mathcal{H}_{i}^2$]
				}}
				{
					$I \gets I \setminus \{i\} $
				} 
			}
		}

		$k \gets k+1$
	}
	\textbf{return} ``\textit{$\Phi$ has been satisfied}''
\end{algorithm}

The following theorem establishes the correctness of Algorithm~1 for Problem~\ref{prob}.

\begin{mythm}\label{thm}\upshape
    Given a dynamical system of form \eqref{eq:system} and an STL formula $\Phi$ as in \eqref{eq:stl-seg}, Algorithm 1 correctly solves Problem~\ref{prob} if all feasible sets in \eqref{eq:feasibleset} are computed exactly, i.e., it is both sound and complete.
\end{mythm}

\vspace{-6pt}

\begin{pf}
If the conclusion of Algorithm~1 is $\mathcal{M}_k = 1$ ``violated prefix", i.e.,  $x_k \notin X_k^I$, then according to Definition~\ref{def:feasibleset}, starting from $x_k$, for any control input  $\mathbf{u}_{k: T-1}$, we have  $x_k \xi_f(x_{k}, \mathbf{u}_{k:T-1}) \not\models \hat{\Phi}_{k}^I$. Therefore, we know that $\forall \mathbf{u}_{k: T-1}:\mathbf{x}_{0:k} \xi_f(x_k, \mathbf{u}_{k:T-1}) \not\models \Phi$,  which is the same as the definition of violated prefix in the beginning of Section \ref{subsec:onlinemonitoring}.
On the other hand, if the conclusion is $\mathcal{M}_k = 0$ ``feasible prefix", i.e.,  $x_k \in X_k^I$, then starting from $x_k$, there exists a control sequence $\mathbf{u}_{k:T-1}$ such that $x_k \xi_f(x_{k}, \mathbf{u}_{k:T-1}) \models \hat{\Phi}_{k}^I$ by Definition~\ref{def:feasibleset}.
Furthermore, all  sub-formulae with index in $\mathcal{I} \setminus I$ have already been satisfied by $\mathbf{x}_{0:k-1}$ for the following reasons.  
From Algorithm~1, we have $\mathcal{M}_{k'} = 0$ and $x_{k'} \!\in\! X_{k'}^{I'}$ for all $k' \!\in\! [0,k-1]$, where $I'$ is the remaining index set at instant $k'$. 
Then for any $i \!\in\! \mathcal{I} \setminus I$, we consider the following two cases:
\vspace{-6pt}
\begin{itemize}
    \item If $O_i = \mathbf{G}$, then we have $X_{k'}^{I'} \subseteq \mathcal{H}_i$ for $k' \!\in\! [a_i, b_i]$ according to the Definition~\ref{def:feasibleset} without specific computation.  Then, $x_{k'} \!\in\! X_{k'}^{I'}$ for all $k' \!\in\! [0,k-1]$ implies that $\mathbf{G}_{[a_i, b_i]} x\!\in\! \mathcal{H}_i$ is satisfied by $\mathbf{x}_{0:k-1}$;
    \medskip
    \item If $O_i = \mathbf{U}'$, assume Line~12 in Algorithm~1 is satisfied and $i$ is removed from $I$ at instant $k'$, then $X_{k''}^{I''} \subseteq \mathcal{H}_i^1$ for $k'' \!\in\! [a_i, k']$ where $I''$ is the remaining index set at instant $k''$ and $i \!\in\! I''$, which implies that $x\!\in\! \mathcal{H}_i^1 \mathbf{U}'_{[a_i, b_i]} x\!\in\! \mathcal{H}_i^2$ is satisfied by $\mathbf{x}_{0:k-1}$.
    \vspace{-6pt}
\end{itemize}
Also, we can prove that sub-formulae $\mathbf{G}_{[a_i, k-1]}x \!\in\! \mathcal{H}_i, i\!\in\! I \cap \mathcal{I}_k$ have been satisfied in the same way.
Hence, the prefix $\mathbf{x}_{0:k}$ is indeed feasible for the entire $\Phi$, since $\hat{\Phi}_{k}^I$ can be satisfied, and sub-formulae with index $\mathcal{I}\setminus I$ and ``Always sub-formulae'' with index $I \cap \mathcal{I}_k$ have been satisfied.
\end{pf}

\begin{remark}\upshape
Theorem~1 states the properties of Algorithm~1 for the ideal case where all feasible sets can be precisely computed. In general, over-/inner-approximation techniques are needed for the computation of feasible sets. If we compute  feasible sets $X_k^I$ by over-approximations, then miss-alarms may be possible since we allow  states that are not actually feasible. On the other hand, if we compute feasible sets by inner-approximations, then false-alarms may be possible.
For safety-critical systems, however, it is more meaningful to use inner-approximations in order to avoid miss-alarms. In terms of our implementation in Section \ref{sec:case}, we adopt  inner-approximations for Cases 1, 3 and 4,  and for Case~2, the  feasible sets are computed exactly due to its simple system dynamic.
\end{remark}


\begin{remark}\upshape
Compared with the direct  approach discussed in Remark~\ref{remark:direct}, the major  advantage of the proposed online monitoring algorithm is  that the online computation burden is very low.  At each time instant, instead of solving a complicated satisfaction problem on-the-fly, our approach just needs to check a set membership. Particularly, the $I$-remaining feasible sets can be computed in an offline fashion and stored in the monitor. Furthermore, our algorithm is only based on the current state $x_k$ and do not need to remember the entire trajectory generated by the system.
\end{remark}

\section{Pre-Computations of Feasible Sets}\label{sec:offline}
In this section, we present methods for the computation of $I$-remaining feasible sets $X_k^{I}$ at time instant $k$.  

\subsection{Computation List}\label{sec:list}
Recall that during online monitoring process, the monitor uses $I$-remaining feasible sets $X_k^{I}$ at each instant $k$, where $I$ is the set of remaining unsatisfied sub-formula index which is determined by state trajectory $\mathbf{x}_{0:k-1}$. 
Since the state trajectory $\mathbf{x}_{0:k-1}$ is unknown a priori at the starting instant of the system, the remaining set $I$ also has multiple possibilities for each instant $k$. 
It seems that we need to compute feasible sets for all subsets in $2^\mathcal{I}$  for each instant $k\in [0,T]$. 
However, the number of remaining sets that are actually possible at each instant is much smaller than the exponential worst-case due to the following observations. 
First, at instant $k$, the remaining set $I$ cannot contain any indices of sub-formulae in $\mathcal{I}_{<k}$ and  that in $\mathcal{I}_{>k}$ are all contained in $I$ since they have not been evaluated.
Second, for each index in $\mathcal{I}_k$, 
if the temporal operator is ``Always'' or $k$ is the first instant of ``Until'', then their index must be in set $I$ since they cannot be satisfied based on the past information.
Formally, we define the \emph{potential index set} for instant $k$ as follows.

\begin{mydef}[Potential Index Set]\upshape 
For each instant $k \!\in\! [0,T]$,  we say subset  $I\subseteq \mathcal{I}$ is a \emph{potential  index set} for instant $k$ if \vspace{-6pt}
\begin{enumerate}[(i)]
    \item 
    $\mathcal{I}_{<k}\cap I=\emptyset$; and\medskip
    \item
     $\mathcal{I}_{>k}\subseteq I$; and\medskip
    \item 
    $\{ i \!\in\! \mathcal{I}_k \mid [O_i=\mathbf{G}] \vee [O_i=\mathbf{U}' \wedge k=a_i]\} \subseteq I$.
\end{enumerate} 
\end{mydef}
We denote by $\mathbb{I}_k$ the set of all potential  index sets for instant $k$. 
Similarly,  we define $\mathbb{X}_k = \{X_k^I \mid I \in \mathbb{I}_k \}$ 
as the set of all \emph{potential feasible sets} for instant $k$. 
Then  our offline objective is to compute all elements in  $\{\mathbb{X}_k:k\in [0,T]\} $.

\begin{myexm}[Cont.]\upshape
	Let us consider the STL formula $\Phi$ in Equation~\eqref{eq:example-prim}.
	For instant $k=7$, the set of all potential   index sets and all potential feasible sets are $\mathbb{I}_7 = \{\{2,3,4,5\}\}$ and $\mathbb{X}_{7} = \{X_{7}^{\{2,3,4,5\}}\}$, respectively.
	For instant $k=8$, we have $\mathbb{I}_{8} = \{ \{4,5\}, \{3,4,5\}\}$ and $\mathbb{X}_{8} = \{X_{8}^{\{4,5\}}, X_{8}^{\{3,4,5\}}\}$.
\end{myexm}

\subsection{Backwards Computation of Feasible Regions}
Now we present our approach for computing all potential feasible sets $\mathbb{X}_k$ for each instant $k$. 
The basis idea is to compute $\mathbb{X}_k$ recursively in a backwards manner. 
Specifically, suppose that we have already known the all potential feasible set in $\mathbb{X}_{k+1}$, 
and then we can use elements in $\mathbb{X}_{k+1}$ to  compute each element in $\mathbb{X}_k$. 

Note that for each remaining set $I \!\in\! \mathbb{I}_k$ for instant $k$, there are multiple choices $I'\!\in\! \mathbb{I}_{k+1}$ for the next instant depending on which currently effective sub-formulae are satisfied.  
Clearly, given the current remaining set $I\!\in\! \mathbb{I}_k$ not arbitrary $I'\!\in\! \mathbb{I}_{k+1}$ can be the remaining set for the next instant. To this end, we introduce the notion of \emph{successor set} as follows.

\begin{mydef}[Successor Sets]\label{def:succ}\upshape
Let $I \!\in\! \mathbb{I}_k$ be a remaining set at instant $k$, 
we say that $I' \!\in\! \mathbb{I}_{k+1}$ is a successor set of $I$ for the next instant if 
\begin{equation}\label{eq:suc}
	\forall i\in \mathcal{I}_{k+1}:    [O_i=\mathbf{U}'\wedge  i\notin I] \Rightarrow i\notin I'.
\end{equation}  
We denote by $\textsf{succ}(I,k) \subseteq \mathbb{I}_{k+1}$ as the set of all successor  sets of $I$ from instant $k$. 
\end{mydef}

Intuitively, $I'$ is a successor set of $I$ says that, for any sub-formula with ``Until" operator, 
if it has been satisfied in $I$, then it should also be satisfied in $I'$.  For the purpose of backwards computation, we define\vspace{-6pt}
\begin{itemize}
    \item 
    $\mathbb{I}_{T+1}=\{\emptyset\}$; and \medskip 
    \item 
    $\forall I\in \mathbb{I}_{T}: \textsf{succ}(I,T)= \mathbb{I}_{T+1}=\{\emptyset\}$. \vspace{-6pt}
\end{itemize}
Note that, the successor set of $I$ may not be unique in general since for those sub-formulae that have not yet been satisfied in $I$ and are still effective at the next instant, they can be either in $I'$ or not depending on the current state of the system.    
To capture this issue,  we define the \emph{satisfaction sets} 
and \emph{consistent regions} as follows. 

\begin{mydef}[Satisfaction Sets and Regions]\label{eq:sat}\upshape 
Let $I \in \mathbb{I}_k$ be a remaining set for instant $k$ and  $I'\in \mathbb{I}_{k+1}$ be a successor set of $I$. \vspace{-6pt}
\begin{itemize}
    \item 
    The \emph{satisfaction set} w.r.t.\ pair $(I,I')$ is defined by
\begin{equation}
\textsf{sat}_\textsf{U}(I,I')
=
\{
i\in I:  O_i=\mathbf{U}'\wedge i\notin I'
\}, \vspace{-6pt}
\end{equation}
which is the set of indices of sub-formulae with ``Until" operator that are in $I$ but not in $I'$.

\medskip
\item 
The \emph{consistent region} w.r.t.\ pair $(I,I')$ is defined  by 
\begin{equation}\label{eq:H}
    H_k(I, I') = \bigcap_{i \in I \cap \mathcal{I}_k} H_i, 
\end{equation}
where	
\begin{align}\label{eq:three-case}
		H_i = 
		\left\{
		\begin{array}{cl}
			\mathcal{H}_i^1 \cap \mathcal{H}_i^2 \ \ \ 
			& \text{if}  \ i \!\in\! \textsf{sat}_\textsf{U}(I,I')\\
			\mathcal{H}_i^1   \setminus \mathcal{H}_i^2  
			& \text{if} \  
			O_i\!=\!\mathbf{U}' \wedge 
			i \!\notin\!   \textsf{sat}_\textsf{U}(I,I')  \\
			\mathcal{H}_i
			& \text{if} \   O_i\!=\!\mathbf{G}.
		\end{array}
		\right.  
	\end{align}     
\end{itemize}

\end{mydef}

The intuitions of the above definitions are as follows. 
Suppose that the current remaining set is $I$ at instant $k$ and becomes $I'$ at instant $k+1$. Then satisfaction set $\textsf{sat}_\textsf{U}(I,I')$ captures the indice set of ``Until" formulae that are satisfied at instant $k$. 
Since $I$ is considered as an element in $\mathbb{I}_k$,   we naturally have $\textsf{sat}_\textsf{U}(I,I') \subseteq \mathcal{I}_k$.
In order to trigger the evolution of the remaining set from $I$ to $I'$, 
at instant $k$, 
the system should be in the consistent region $H_k(I, I')$. 
Specifically, we consider each sub-formula  that is remaining and effective at instant $k$, i.e., $i\in I\cap \mathcal{I}_k$. Then we have the following three cases as shown in Equation~\eqref{eq:three-case}:\vspace{-6pt}
\begin{itemize}
    \item 
    If $i \!\in\! \textsf{sat}_\textsf{U}(I,I')$, then it means that the system must be in region $\mathcal{H}_i^1 \cap \mathcal{H}_i^2$ in order to satisfy sub-formula $i$; \medskip
    \item 
    If $O_i\!=\!\mathbf{U}'$ but $i \!\notin\! \textsf{sat}_\textsf{U}(I,I')$, then it means that the
    sub-formula  $i$ is not yet satisfied, and, therefore, the system should be in region $\mathcal{H}_i^1   \setminus \mathcal{H}_i^2$;  \medskip
    \item 
    If $O_i\!=\!\mathbf{G}$, then the system should  stay in region $\mathcal{H}_i$.\vspace{-6pt}
\end{itemize}
Since each sub-formula $i\in I\cap \mathcal{I}_k$ should satisfy the above requirements, $H_k(I, I')$ is taken as the intersection of the region of each sub-formula. 

Now,  for some $I\in \mathbb{I}_k$ and its successor set  $I'\in \mathbb{I}_{k+1}$, 
the computation of feasible set $X_k^I$ can be divided into two parts:\vspace{-6pt}
\begin{itemize}
	\item 
	First, it should stay in the consistent region $H_k(I, I')$ at instant $k$;
	\medskip
	\item 
	Also, it needs to be able to reach region $X_{k+1}^{I'}$ in one step to satisfy the subsequent requirements.\vspace{-6pt}
\end{itemize}
This observation is formalized with the help of \emph{one-step feasible set} defined as follows. 

\begin{mydef}[One-Step Feasible Set]\upshape
Let $\mathcal{S}\subseteq \mathcal{X}$ be a set of states representing the ``target region". 	Then the one-step feasible set of $\mathcal{S}$ is defined by 
		\begin{equation}\label{eq:one-step-f}
			\Upsilon(\mathcal{S}) = \{x \in \mathcal{X} \mid \exists u \in \mathcal{U}  \text{ s.t. }  f(x, u) \in \mathcal{S}\}. 
		\end{equation}
\end{mydef}

In terms of our computation of feasible regions, 
if the system is evolving from $I$ to $I'$ and maintains the satisfiability of $I'$ from instant $k+1$, then we know that the system should be in   region $H(I, I') \cap \Upsilon(X_{k+1}^{I'})$ at instant $k$. 
However, the remaining set $I'$ for the next instant depends on the current-state of the system. Therefore, to compute $X^I_k$, we need to consider all possible successor sets $I' \!\in\! \textsf{succ}(I,k)$, and take the union of these regions. This is formalized by the following theorem.

\begin{mythm}\upshape
	Suppose that $I$ is the remaining index set and $\mathbb{X}_{k+1}$ is the set of all potential feasible sets at next time instant.
	Then $I$-remaining feasible set $X_k^I$ defined in Definition~\ref{def:feasibleset} for the time instant $k$ can be computed as follows
	\begin{equation}\label{eq:dynpro}
		X_k^I = \bigcup_{I' \in \textsf{succ}(I, k)} \Big( H_k(I, I') \cap \Upsilon(X_{k+1}^{I'}) \Big).
	\end{equation} 
\end{mythm}

\begin{pf}
\begin{figure*} \tiny
	\begin{align}\label{eq:proof}
		& \hat{\Phi}_{k}^{I} 
		= \bigwedge_{i   \in I\cap \mathcal{I}_{k}   } \Phi_i^{[k,b_i]} \wedge \bigwedge_{i \in \mathcal{I}_{>k}   } \Phi_i^{[a_i,b_i]}
		= \bigwedge_{i \in I_k \setminus \mathcal{I}_{k+1}} \Phi_i^{[k,k]} \wedge \underline{\bigwedge_{i \in I_k \cap \mathcal{I}_{k+1}} \Phi_i^{[k,b_i]}}
		\wedge \bigwedge_{i \in \mathcal{I}_{>k}} \Phi_i^{[a_i,b_i]} 
		\\
		& = \bigwedge_{i \in I_k \setminus \mathcal{I}_{k+1}} \Phi_i^{[k,k]} 
		\underline{
		\dashuline{
		\wedge \bigwedge_{i \in I_k^\mathbf{U} \cap \mathcal{I}_{k+1}} \Phi_i^{[k,b_i]}}
		\wedge \bigwedge_{i \in I_k^\mathbf{G} \cap \mathcal{I}_{k+1}} \Phi_i^{[k,b_i]}}
		\wedge \bigwedge_{i \in \mathcal{I}_{>k}} \Phi_i^{[a_i,b_i]} \nonumber \\
		& = \dashuline{\bigvee_{\hat{I} \subseteq I_k^\mathbf{U} \cap \mathcal{I}_{k+1}} }
		\Big(
			\underbrace{
			\bigwedge_{i \in I_k \setminus \mathcal{I}_{k+1}} \Phi_i^{[k,k]} 
			\dashuline{
			\wedge \bigwedge_{i \in I_k^\mathbf{U} \cap \mathcal{I}_{k+1} \setminus \hat{I}} \Phi_i^{[k,k]}
			\wedge \bigwedge_{i \in \hat{I}} \mathbf{G}_{[k,k]}x \!\in\! \mathcal{H}_i^1
			}
			\wedge \bigwedge_{i \in I_k^\mathbf{G} \cap \mathcal{I}_{k+1}} (\Phi_i^{[k,k]} 
			}_{\psi_1(\hat{I})}
			\wedge
			\underbrace{
			\Phi_i^{[k+1,b_i]})
			\dashuline{
			\wedge \bigwedge_{i \in \hat{I}} \Phi_i^{[k+1,b_i]}
			}
			\wedge \bigwedge_{i \in \mathcal{I}_{>k}} \Phi_i^{[a_i,b_i]}
			}_{\psi_2(\hat{I})}
		\Big)  \nonumber
	\end{align}
\end{figure*}
    When $k = T$, since $\textsf{succ}(I, k) = \{\emptyset\}$ and $\textsf{sat}_\textsf{U}(I,I') = \textsf{sat}_\textsf{U}(I,\emptyset)  = \{i \!\in\! I: O_i = \mathbf{U}'\}$, from Equation~\eqref{eq:dynpro}, we have
    \[
    X_T^I = \bigcap_{i \in \textsf{sat}_\textsf{U}(I,I')} (\mathcal{H}_i^1 \cap \mathcal{H}_i^2) \cap \bigcap_{i \in I \setminus \textsf{sat}_\textsf{U}(I,I')} \mathcal{H}_i,
    \]
    which is clearly the $I$-remaining feasible set of $\hat{\Phi}_{k}^{I} = \bigwedge_{i \in I} \Phi_i^{[T,T]}$.

	For the case of $k \neq T$, we can write 
	$\hat{\Phi}_{k}^{I}$ in the form of Equation~\eqref{eq:proof},	where $I_k$ is the abbreviation of $I\cap \mathcal{I}_k$, and $I_k^\mathbf{G}, I_k^\mathbf{U}$ are the sets of ``Always'' and ``Until'' indices contained in $I_k$, respectively.
	Intuitively,   Equation~\eqref{eq:proof}  divides $I_k$ into two parts, $I_k\setminus \mathcal{I}_{k+1}$ and $I_k\cap \mathcal{I}_{k+1}$, which are the  sets of indices of the last instants and the non-last instants, respectively. 
	We can further split $I_k\cap \mathcal{I}_{k+1}$ according to different temporal operators as shown in the second line of the Equation~\eqref{eq:proof}. 
	For ``Until'' sub-formulae with indices in $I_k^\mathbf{U} \cap \mathcal{I}_{k+1}$, it can be  either satisfied currently or postponed to next instant. 
	We denote by $\hat{I}$ the index set for those sub-formulae that are not satisfied currently, which can be any subset of $I_k^\mathbf{U}\cap \mathcal{I}_{k+1}$.  
	Then $\hat{\Phi}_{k}^{I}$ can be further rewritten as the third line of  Equation~\eqref{eq:proof}, 
	where for each possible $\hat{I}$, we divide the corresponding formula into two parts: 
	one only related to the current state (denoted by $\psi_1(\hat{I})$) and the other related to the future requirements (denoted by $\psi_2(\hat{I})$).

   We observe that, for any  $\hat{I}$, we have $x_{k}\xi(x_{k}, \mathbf{u}_{k:T-1}) \models \psi_1(\hat{I}) \wedge \psi_2(\hat{I})$ if the following holds: \vspace{-6pt}
	\begin{itemize}
		\item $x_k \models \psi_1(\hat{I})$,
		\medskip
		\item $x_{k+1}\xi(x_{k+1}, \mathbf{u}_{k+1:T-1}) \models \psi_2(\hat{I})$,
		\medskip
		\item $x_{k+1}$ is reachable from $x_k$ under some $u_k$.\vspace{-6pt}
	\end{itemize}
	The first condition holds iff  $x_k$ stays in region $H_k(I, I')$ with $I' = \mathcal{I}_{>k} \cup (I_k^\mathbf{G} \cap \mathcal{I}_{k+1}) \cup \hat{I} $ and $\textsf{sat}_\textsf{U}(I, I') = (I_k^\mathbf{U} \setminus \mathcal{I}_{k+1}) \cup (I_k^\mathbf{U} \cap \mathcal{I}_{k+1} \setminus \hat{I})$. 
	The second condition holds iff  $x_{k+1}$ is in $I'$-remaining feasible set $X_{k+1}^{I'}$. 
	The third condition holds iff  $x_k \!\in\! \Upsilon(X_{k+1}^{I'})$. 
	Therefore,  $\psi(\hat{I}):=\psi_1(\hat{I}) \wedge \psi_2(\hat{I})$ holds iff  $x_k$ is in $H_k(I, I') \cap \Upsilon(X_{k+1}^{I'})$. 
	Finally, recall that $\hat{\Phi}_{k}^{I}$ is the disjunction of all possible $\psi(\hat{I})$. This suffices to consider all possible $I' \in \textsf{succ}(I,k)$.
	Therefore, we have $X_k^I = \bigcup_{I' \in \textsf{succ}(I, k)} \Big( H_k(I, I') \cap \Upsilon(X_{k+1}^{I'}) \Big)$ which is the same as Equation~\eqref{eq:dynpro}, i.e., the theorem is proved.	
\end{pf}

%

\begin{myexm}
	[Cont.]\upshape
	Let us consider the STL formula $\Phi$ in Equation~\eqref{eq:example-prim}.
	For instant $k=11$, assume that remaining set is $I = \{3, 4, 5\} \in \mathbb{I}_{11}$.
	In this case, any set in $\mathbb{I}_{12} = \{\emptyset, \{3\}, \{5\}, \{3,5\}\}$ could be a possible successor set as defined in Definition~\ref{def:succ}, i.e., $\textsf{succ}(I, 11) = \mathbb{I}_{12}$, and we denote by $I'_1 = \emptyset$,  $I'_2 = \{3\}$, $ I'_3 = \{5\}$ and $ I'_4 = \{3, 5\}$.
	Note that for sub-formula $\Phi_3 = \mathbf{F}_{[5,15]} x \!\in\! \mathcal{H}_\mathbf{F}$, we write it as $x \!\in\! \mathcal{X} \mathbf{U}'_{[5,15]} x \!\in\! \mathcal{H}_\mathbf{F}$ here.
	For four possible successor sets, their satisfaction sets are \vspace{-6pt}
	\begin{align}
		& \textsf{sat}_\textsf{U}(I, I'_1) = \{3, 5\}, \textsf{sat}_\textsf{U}(I, I'_2) = \{5\}, \nonumber\\ 
		& \textsf{sat}_\textsf{U}(I, I'_3) = \{3\}  \ \text{and} \  \textsf{sat}_\textsf{U}(I, I'_4) = \emptyset \nonumber
	\end{align}
	respectively, and their consistent regions are\vspace{-6pt}
	\begin{align}
		H_k(I, I'_1) =& (\mathcal{H}_\mathbf{U}^1 \cap \mathcal{H}_\mathbf{U}^2) \cap (\mathcal{X} \cap \mathcal{H}_\mathbf{F}) \cap \mathcal{H}_\mathbf{G},\nonumber \\
	    H_k(I, I'_2) =& (\mathcal{H}_\mathbf{U}^1 \cap \mathcal{H}_\mathbf{U}^2) \cap (\mathcal{X} \setminus \mathcal{H}_\mathbf{F}) \cap \mathcal{H}_\mathbf{G}, \nonumber\\
	    H_k(I, I'_3) =& (\mathcal{X} \cap \mathcal{H}_\mathbf{F}) \cap (\mathcal{H}_\mathbf{U}^1 \setminus \mathcal{H}_\mathbf{U}^2) \cap \mathcal{H}_\mathbf{G},\nonumber \\
		H_k(I, I'_4) =& (\mathcal{H}_\mathbf{U}^1 \setminus \mathcal{H}_\mathbf{U}^2) \cap (\mathcal{X} \setminus \mathcal{H}_\mathbf{F}) \cap \mathcal{H}_\mathbf{G},\nonumber
		\vspace{-6pt}
	\end{align} 
  respectively.
	Then, we have $I$-remaining feasible set at instant $11$ is \vspace{-6pt}
	\begin{align}
		X_{11}^I = 
		& \big(H_k(I, I_1') \cap \Upsilon(X_{12}^{I'_1}) \big) \cup \big(H_k(I, I_2') \cap \Upsilon(X_{12}^{I'_2}) \big) \nonumber \\
		& \cup \big(H_k(I, I_3') \cap \Upsilon(X_{12}^{I'_3}) \big) \cup \big(H_k(I, I_4') \cap \Upsilon(X_{12}^{I'_4}) \big). \nonumber
	\end{align}
\end{myexm}

\subsection{Offline Computation Algorithm}
\begin{algorithm}[ht]
	\caption{Offline Computations of Feasible Sets}
	\KwIn{STL formula $\Phi = \bigwedge_{i=1}^{N} \Phi^{[a_i, b_i]}$}
	\KwOut{All potential sets $\{\mathbb{X}_k:k\in [0,T]\}$} 
	for each $k \in[0,T]$, $\mathbb{X}_k \gets \emptyset$   \\
	$X_{T+1}^\emptyset \gets \mathbb{R}^n$ \\
	$k \gets T$ \\
	\While{$k \geq 0$}
	{
		\ForAll{$I \in \mathbb{I}_k$}
		{
			$X_k^I \gets \bigcup_{I' \in \textsf{succ}(I, k)} \Big( H_k(I, I') \cap \Upsilon(X_{k+1}^{I'}) \Big)$ \\
			$\mathbb{X}_k \gets \mathbb{X}_k \cup \{X_k^I\}$  
		}
		$k \gets k-1$
	}
\end{algorithm}

Now, we present  the complete procedure for offline computations of feasible sets in Algorithm~2.  
Initially, for each instant $k\in [0,T]$, we set $\mathbb{X}_k$ as the empty set since no $X_k^I$ is computed for some $I\in \mathbb{I}_k$ (line~1). 
Then for the unique element $X_{T+1}^\emptyset$ in $\mathbb{X}_{T+1}$, we set it as $\mathbb{R}^n$ since the formula has already been finished at instant $T$ and there is no requirement at instant $T+1$.  
To proceed the backwards induction, 
at each instant $k\in [0,T]$, for each $I\in \mathbb{I}_k$, we compute $X_k^I$ according to Equation~\eqref{eq:dynpro} by considering its all possible successor sets at instant $k+1$ (lines 4-8).

\subsection{Numerical Computation Considerations}\label{sec:appro}
Finally, we conclude this section by discussing some considerations in the numerical computation of above feasible sets.  

\textbf{Computations of One-Step Sets:}
In order to realize Algorithm~2, the key is to compute one-step feasible set $\Upsilon(\cdot)$. In general, there is no close-form expression for such sets and the computation highly depends on the dynamic of the system.  Particularly, (inner or outer) approximation methodologies have been widely used in practice to achieve the trade off between the  computational  accuracy and complexity.
For example, for linear systems, computation methods for one-step set have been presented subject to polytopic constraints described by linear differential inclusions or for piece-wise affine systems; see, e.g.,  \cite{blanchini1994ultimate, kerrigan2001robust, mayne2001control}. For general nonlinear systems, however, computing the one-step set precisely is much more challenging \cite{stipanovic2003computation, mitchell2003overapproximating, bravo2005computation}. For example, \cite{bravo2005computation} proposed a branch and bound algorithm with interval arithmetic approach which provides an inner approximation with a given bound of the error. Furthermore, we can also use some data-driven methods to compute it, and similar  framework can be found in \cite{devonport2020data}.

In our implementation in Section \ref{sec:case}, we adopt the generic method in \cite{bravo2005computation}. The main idea of \cite{bravo2005computation} is to start with a large interval box that contains all feasible (satisfiable) sets for sure and then to iteratively divide the box into smaller boxes and track which boxes are still within the feasible (satisfiable) sets by some forward reachable analysis tools; the readers are referred to \cite{bravo2005computation} for more details. 
Note that the original algorithm in \cite{bravo2005computation} supports inner-approximation and can be easily extended to over-approximation.


\textbf{Complexity of Pre-Computations:}
To perform the pre-computations, first, we need to  compute region $\mathcal{H}^\mu$ for each  $\pi^\mu$. This complexity generally depends on the nonlinearity degree of the predication function  \cite{evtushenko2018approximating}. 
The complexity for computing all feasible sets is linear in the horizon of the entire formula. 
Also, since we just need to consider sets in $\mathbb{I}_k$ for each instant rather than the entire power set of the index set, the total number of feasible sets  to be computed generally will not grow exponentially as the horizon increases.
However, the complexity of computing feasible sets at instant $k$ is exponential in terms of the cardinality of index set $\mathcal{I}_k$, especially the cardinality of index of ``Until'' operators in $\mathcal{I}_k$, since the cardinality of $\mathbb{I}_k$ will explode exponentially as the number of ``Until'' operators in $\mathcal{I}_k$ grows.
For each step in the iteration,  the complexity for computing the one-step sets for constrained systems largely depends on the system model and may increase exponentially with the order of the system. 
Finally, it is worth  mentioning again that the computations of feasible sets are purely offline, which do  not affect the complexity of the online execution of the monitoring algorithm.



\section{Extension to the Case  of Satisfied Prefixes}\label{sec:extension}
In the previous sections, we have formulated and solved the problem to detect violated prefixes. 
In some scenarios, however, \emph{satisfied prefixes} are also of interest  since the monitor may claim the satisfaction of task in advance and save   resources for future processes.
Formally, we say a prefix signal $\mathbf{x}_{0:k}$ is \emph{satisfied} if for any control input $\mathbf{u}_{k: T-1}$, we have $\mathbf{x}_{0:k} \xi_f(x_k, \mathbf{u}_{k:T-1}) \models \Phi$. 
Our framework can be easily extended to the case where one is  also interested in detecting satisfied prefix before the task is actually  satisfied. 
Here,  we introduce this extension briefly. 

\textbf{Online Monitoring:}
To detect satisfied prefixes online, we just need to  check whether or not the system can always fulfill those $I$-remaining sub-formulae  that have not been achieved up to now in the future.
To capture this information, we introduce  notion similar to Definition~\ref{def:feasibleset}, called $I$-remaining satisfiable set as follows: 
\begin{align}\label{eq:satisset}
	Y_k^{I} \! =\! 
	\left\{ 
	x_k \in \mathcal{X} \,\middle\vert\, \!\!\!\!
	\begin{array}{cc}
		\forall \ \mathbf{u}_{k:T-1} \in \mathcal{U}^{T-k} \\
		\text{ s.t. } x_k \xi_f(x_{k}, \mathbf{u}_{k:T-1}) \models \hat{\Phi}_{k}^I
	\end{array} 
	\right\}. \nonumber
\end{align}
The only difference between the $I$-remaining feasible set and the satisfiable set is their quantifiers ``$\exists$'' and ``$\forall$''. Clearly, we have  $Y_k^{I} \subseteq X_k^{I}$.
Then for online monitoring Algorithm~1, 
to detect satisfied prefixes, one can just add a new testing condition after line~9: 
if $x_k \!\in\! Y_k^{I}$, then return ``\textit{prefix is satisfied}''.
The soundness and completeness analysis of this case is consistent with that of monitoring algorithm for only feasible prefixes in Theorem \ref{thm}.

\textbf{Offline Computation:}
The computation method for satisfiable set $Y_k^{I}$ is also similar to the case of feasible set and can be done with the help of one-step satisfiable set as follows:
\begin{equation}\label{eq:one-step-s}
	\hat{\Upsilon}(\mathcal{S}) = \{x \in \mathcal{X} \mid \forall u \in \mathcal{U}  \text{ s.t. }  f(x, u) \in \mathcal{S}\}. \nonumber
\end{equation}
Correspondingly, the $I$-remaining satisfiable set $Y_k^I$ can be computed with Equation~\eqref{eq:dynpro} by replacing one-step feasible set $\Upsilon(\cdot)$ to satisfiable set $\hat{\Upsilon}(\cdot)$ as follows:
\begin{equation}
	Y_k^I = \bigcup_{I' \in \textsf{succ}(I, k)} \Big( H_k(I, I') \cap \hat{\Upsilon}(X_{k+1}^{I'}) \Big), \nonumber
\end{equation} 
and for offline computation we just need to repeat lines~6-7 again after it in the form of satisfiable set.
The correctness can be also established in the same way which are omitted here.

\section{Case Studies for Online Monitoring}\label{sec:case}

We have implemented our   proposed  computation methods for feasible sets as well as satisfiable sets in \textsf{Julia} language \cite{bezanson2017julia}  with the help of existing package \textsf{JuliaReach} \cite{bogomolov2019juliareach, lazysets21, benet2019taylormodels}.  
All algorithms are carried out by a  computer with i9-9900K CPU 3.60GHz and 32 GB of RAM.
Our codes are available at  \url{https://github.com/Xinyi-Yu/MPM4STL}, where more details on the computations can be found.

Then  we illustrate our online monitoring algorithm by applying it to  four different case studies: building temperature control, double integrator, nonholonomic mobile robot and spacecraft rendezvous. 
Specifically, the offline computations are performed with one-step set computation method in  \cite{bravo2005computation} with some approximations except the second case study, 
whose feasible sets are computed analytically and exactly due to its  simple linear model structure.
Furthermore, for the first and the third case studies, we detect both violated and satisfied prefixes, while for the second and the fourth case studies, we only detect violated prefixed since most of their $I$-remaining satisfiable sets are empty. 
The total execution times for pre-computations of feasible sets in the four case studies are 1 minute, 55 minutes, 3.3 hours, and 2.1 hours, respectively. The shorter execution time for the first case study can be attributed to its simpler system dynamics compared to the other case studies.
We show that,  by leveraging the model information of the dynamical system, our model-based approach may provide better monitoring evaluations compared with purely model-free approaches.

\subsection{Building Temperature Control}

We consider the problem of monitoring the temperature of a single zone building whose dynamic is given by the following difference equation
\[
	x_{k+1} = x_k + \tau_s(\alpha_e(T_e - x_k) + \alpha_H(T_h - x_k )u_k),
\]
where state $x_k \!\in\! \mathcal{X}=[0,45]$ denotes the zone temperature of building the at instant $k$, control input $u_k \!\in\! \mathcal{U} = [0,1]$ is the ratio of the heater valve, $\tau_s = 1$ minute is the sampling time, $T_h = 55^\circ C$ is the heater temperature, $T_e = 0^\circ C$ is outside temperature, and $\alpha_e = 0.06$ and $\alpha_H = 0.08$ are the heat exchange coefficients. The model is adopted from \cite{jagtap2020formal}.

\begin{figure}[t]
	\centering
	\includegraphics[width = 230pt]{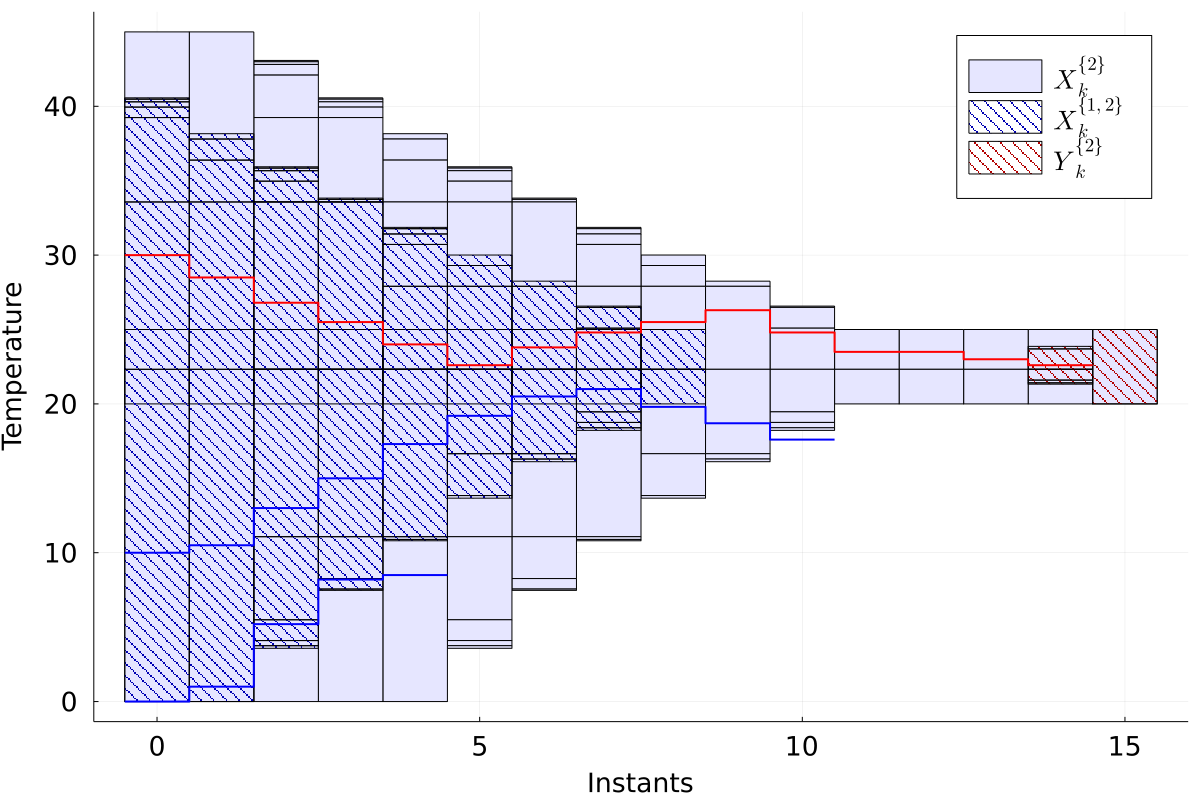}
	\caption{Three  signals for the building temperature control system.}
	\label{fig:case1}
\end{figure}

The objective of the temperature control  system is to warm the single room to specific comfortable environment between $20^\circ C - 25^\circ C$  in $8$ minutes, and  then keep the temperature in this interval from $10$ to $15$ minutes, which can be described by the following STL formula 
\[
	\Phi = \mathbf{F}_{[0,8]} x_k \!\in\! [20,25] \wedge \mathbf{G}_{[10,15]} x_k \!\in\! [20,25].
\]
Before starting the online monitoring process,  we first compute the $I$-remaining  feasible and satisfiable sets of STL formula $\Phi$ by Algorithm~2 for each time instant $k$ and for all $I \!\in\! \mathbb{I}_k$. 
The offline computation results are shown in Fig.~\ref{fig:case1}. 
Specifically, areas filled with black dotted lines and blue colors are $\{1,2\}$-remaining feasible set and $\{2\}$-remaining feasible set respectively, 
and red dotted lines denote  $\{2\}$-remaining satisfiable sets.
Some black horizontal lines in the figure was caused by one-step set computation since the computation needs to divide boxes into smaller ones.
Note that  satisfied sets $Y_k^{\{1,2\}}$ for all instants and $Y_k^{\{2\}}$ for $k \!\in\! [1, 13]$ are all empty.

During the online monitoring process, the monitor observes the current state at each time and makes evaluations. For example, let us consider three possible state traces generated by the system shown as the two blue lines and a red line  in Fig.~\ref{fig:case1}.
At instant $k=4$ and $k=10$ respectively,  using the model-free approach, one can only make the inconclusive evaluations for the two blue lines since the future signals can either satisfy remaining tasks or not without any constraints. 
However, using our model-based approach, since $x_{4} \!\notin\! X_{4}^{\{1,2\}}$ for the below blue line and $x_{10} \!\notin\! X_{10}^{\{2\}}$ for the above, 
we can conclude immediately that the formula will be violated inevitably since there exists no controller under which the STL formula can be satisfied.
Therefore, compared with existing model-free algorithms \cite{ho2014online, deshmukh2017robust}, our method can claim the violation of specification in advance at instant $4$ and $10$ respectively, while existing algorithms cannot provide a clear violation conclusion.
Also, for the red line, we can claim at instant $14$ that the task will be satisfied definitely in the future since $x_{14} \!\in\! Y_{14}^{\{2\}}$ which implies that the next state $x_{15}$ will be in $[20,25]$ no matter what controller $u_{14}$ is, although it has not been satisfied currently according to  model-free approaches.

\subsection{Double Integrator}

\begin{figure}[t]
	\centering
	\includegraphics[width = 230pt]{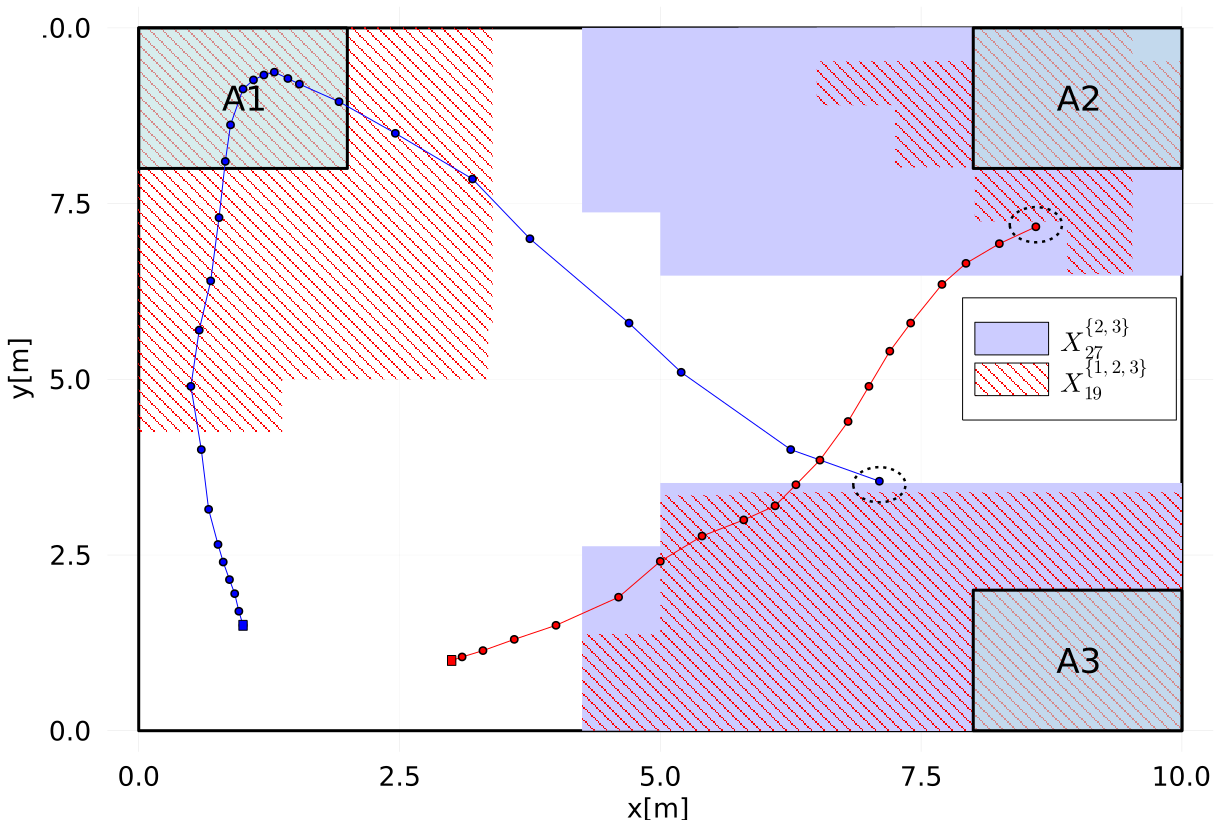}
	\caption{Two trajectories for the double integrator system.}
	\label{fig:case2}
\end{figure}

For the second case study, we consider the planar motion of a single robot with double integrator dynamic, where the system model and temporal logic task are  similar to those studied in \cite{lindemann2017robust}. The  system model with a sampling period of $0.5$ seconds is as follows 
\[
	x_{k+1} = 
	\begin{bmatrix}
		1 & 0.5 & 0 & 0 \\
		0 & 1 &  0 & 0 \\
		0 & 0 & 1 & 0.5 \\
		0 & 0 & 0 & 1 
	\end{bmatrix}
	x_{k} + 
	\begin{bmatrix} 
		0.125 & 0   \\ 
		0.5 & 0 \\ 
		0 & 0.125 \\ 
		0 & 0.5 
	\end{bmatrix}
	u_k,
\]
where state $x_k = [x \ v_x \ y \ v_y]^T$ denotes $x$-position, $x$-velocity, $y$-position and $y$-velocity, and control input $u_k = [u_x \ u_y]^T$ denotes $x$-accelation and $y$-accelation, respectively.
The physical constraints are $x \!\in\! \mathcal{X} = [0, 10] \times [-1.5, 1.5] \times [0,10] \times [-1.5, 1.5]$ and $u \!\in\! \mathcal{U} = [-1, 1] \times [-1,1]$.

The objective of the robot is to visit all three regions A1, A2 and A3   shown in Fig.~\ref{fig:case2} within  time interval between 10 to 40 instants (i.e., $5$ to $20$ seconds) in any order, which can be described by the following STL formula 
\[
	\Phi = \Phi_1 \wedge \Phi_2 \wedge \Phi_3, 
\]
where $\Phi_1 = \mathbf{F}_{[10,40]}(x \!\in\! [0,2] \wedge y \!\in\! [8,10])$, $\Phi_2 = \mathbf{F}_{[10,40]}(x \!\in\! [8,10] \wedge y \!\in\! [8,10])$ and $\Phi_3 = \mathbf{F}_{[10,40]}(x \!\in\! [8,10] \wedge y \!\in\! [0,2])$.

We consider two trajectories of the robot starting from two rectangle points up to instant $k = 27$ and $k = 19$ (i.e., 13.5 seconds and 9.5 seconds) shown in Fig.~\ref{fig:case2}, respectively, where the feasible sets $X_{27}^{\{2,3\}}$ and $X_{19}^{\{1,2,3\}}$ computed offline are also depicted. 
For the blue trajectory, at instant $k=12$ the first time robot comes to A1, the remaining index set turns from $I=\{1,2,3\}$ to $\{2,3\}$ and then we use $X_k^{\{2,3\}}$ to monitor. At instant $27$, we observe that $x_{27} \notin X_{27}^{\{2,3\}}$ and the monitor can claim that the robot will never satisfy the task. Similarly, for the red trajectory, it does not visit any region of interest before $k =19$ and also $x_{19} \notin X_{19}^{\{1,2,3\}}$. Then we can claim in advance that it will not complete the task.
Note that the offline results shown in the figure is the projection to the first and third dimensions but the set membership   is still checked in the complete 4-dimensional state space (also for later case studies).

\subsection{Nonholonomic Mobile Robot}
\begin{figure}[t]
	\centering
	\includegraphics[width = 230pt]{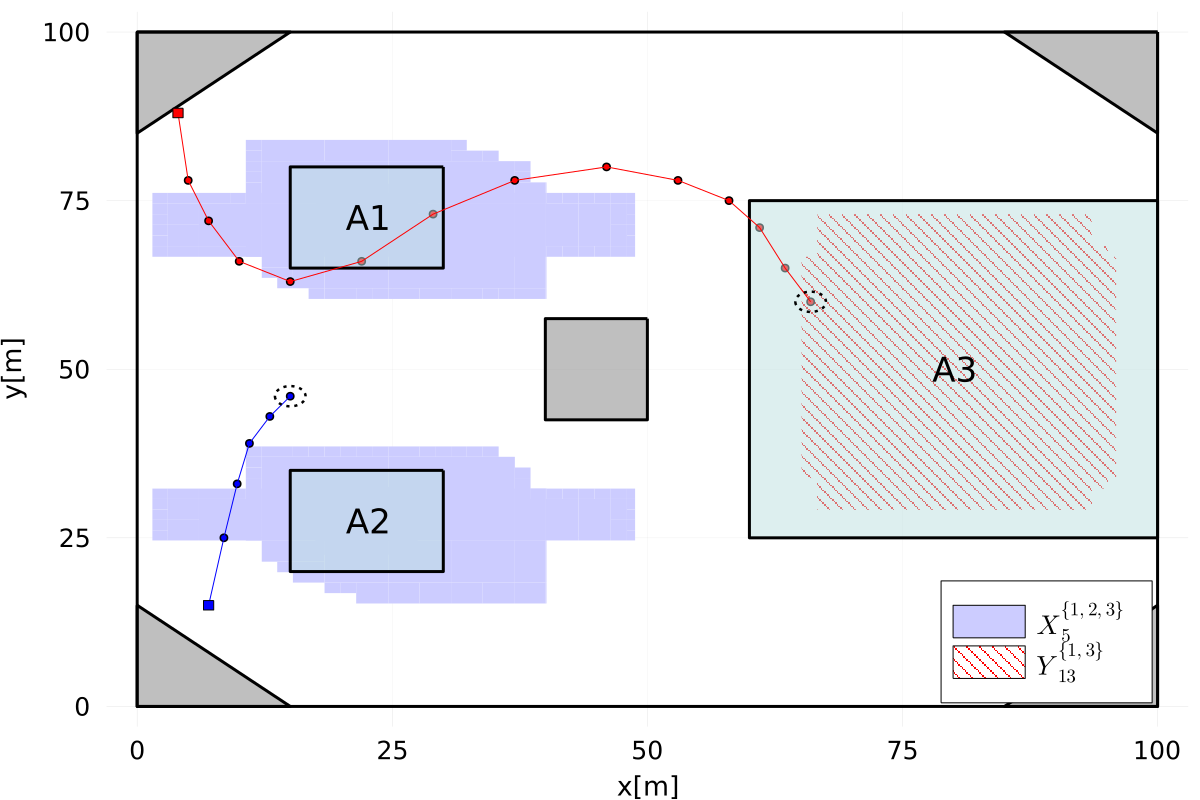}
	\caption{Two   trajectories for the nonholonomic mobile robot.}
	\label{fig:case3}
\end{figure}

Consider a nonholonomic mobile robots modeled by kinematic unicycles \cite{xiao2022decentralized} in the form of
\[
	\dot{x} = v \cos \theta, \
	\dot{y} = v \sin \theta, \
	\dot{\theta} = \omega,
\]
and we discretize it with sampling time 0.5 seconds where state  $x_k = [x \ y \ \theta]^T$ denotes $x$-position, $y$-position and angle, and control input $u_k = [v \ \omega]^T$ denotes speed and angular velocity, respectively.
The physical constraints are $x \!\in\! \mathcal{X} = [0, 100] \times [0, 100] \times [-\frac{\pi}{2}, \frac{\pi}{2}]$ and $u \!\in\! \mathcal{U} = [-10, 10] \times [-0.3, 0.3]$ if $k \leq 10$ and $\mathcal{U} = [-3, 3] \times [-0.3, 0.3]$ otherwise.

The workspace of the mobile robot is shown in Fig.~\ref{fig:case3}.
The objective is to first visit regions A1 or A2 (green)  with a specific angle before instant 8 (i.e., 4 seconds) and then to stay in A3 (green) between instants 10 and 15 (i.e., 5 seconds and 7.5 seconds). Meanwhile, the robot should always avoid the obstacles (grey) in the map. The task can be described by the following STL formula
\[
	\Phi = \Phi_1 \wedge \Phi_2 \wedge \Phi_3,
\]
where $\Phi_1 = \mathbf{G}_{[0,15]} (y < x+85 \wedge y>-x+15 \wedge y<-x+185 \wedge y>x-85 \wedge \neg (x \!\in\! [40,50] \wedge y \!\in\! [42.5,47.5]))$, 
$\Phi_2 = \mathbf{F}_{[0,8]} ((x \!\in\! [15,30] \wedge y \!\in\! [20,35] \wedge \theta \!\in\! [-1.4, -0.2]) \vee (x \!\in\! [15,30] \wedge y \!\in\! [65,80] \wedge \theta \!\in\! [-1.4, -0.2]))$ and 
$\Phi_3 = \mathbf{G}_{[10,15]} (x \!\in\! [60, 100] \wedge y \!\in\! [25,75])$.

We consider two trajectories of the robot starting from two rectangle points up to $k=5$ in blue and up to $k=13$ in red as shown in Fig.~\ref{fig:case3}, respectively. 
The monitor can claim that the blue trajectory will violate the task at $k=5$ since $x_5^{blue} \!\notin\! X_5^{\{1,2,3\}}$. 
Also, we can claim that  red trajectory will  satisfy the specification for sure at $k=13$ since $x_{13}^{red} \!\in\! Y_{13}^{\{1,3\}}$.

\subsection{Spacecraft Rendezvous}

\begin{figure}[t]
	\centering
	\includegraphics[width = 230pt]{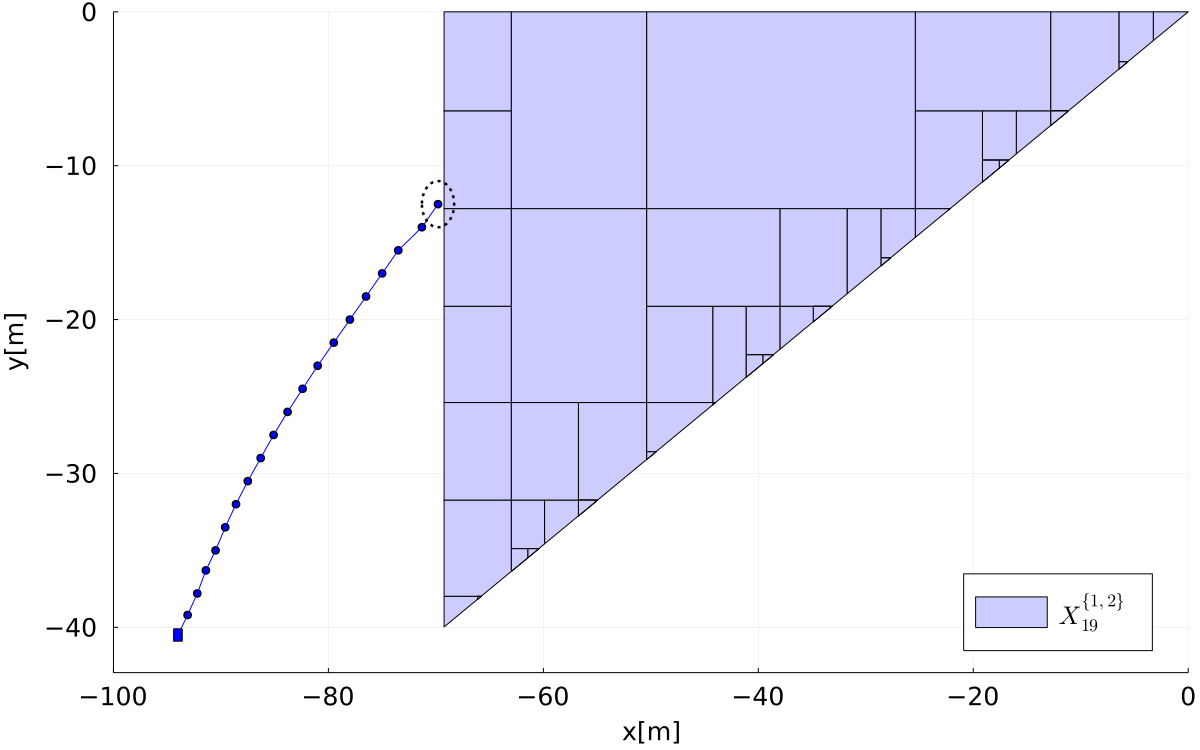}
	\caption{One possible signal for spacecraft rendezvous case.}
	\label{fig:case4}
\end{figure}

Finally, we  consider a spacecraft rendezvous example  adopted from \cite{chan2017verifying}.
The model is the so-called Hill's relative coordinate frame centered on the \emph{target} spacecraft, and the following nonlinear dynamic equations describe the two-dimensional, planar motion of \emph{chaser} spacecraft on an orbital plane towards a target spacecraft
\begin{align}
	&\dot{x} = v_x,  \nonumber \\
	&\dot{y} = v_y, \nonumber\\
	&\dot{v}_x = n^2x + 2nv_y+ \frac{\mu}{r^2} - \frac{\mu}{r^3_c}(r+x) + \frac{u_x}{m_c},   \nonumber\\
	&\dot{v}_y = n^2y -2nv_x - \frac{\mu}{r^3_c}y + \frac{u_y}{m_c}, \nonumber
\end{align}
where state is $x_k = [x \ y \ v_x \ v_y]^T$, control input is chaser's thrusters $u_k = [u_x \ u_y]^T$ and we discretize the model with sampling time 0.5 minutes. The parameters are $\mu = 3.986 \times 10^{14} \times 60^2 [m^3/min^2], r = 42164\times 10^3 [m], m_c = 500[kg], n = \sqrt{\frac{\mu}{r^3}}$ and $r_c = \sqrt{(r+x)^2+y^2}$. 
When the target and the chaser spacecrafts' separation distance is less than $100m$, the chaser needs to continue to rendezvous and be docked to the target spacecraft with specific angle of approach and closing velocity. Therefore, we focus on monitoring this critical rendezvous period.
Specifically, the $x$-velocity and $y$-velocity should be less than $3.5 [m/min]$ and the thrusters cannot provide more than $10N$ of force in any single direction in this period, i.e., the physical constraints are $x \!\in\! \mathcal{X} = [-100, 0] \times [-100, 0] \times [0, 3.5] \times [0, 3.5]$ and $u \!\in\! \mathcal{U} = [0,10] \times [0,10]$.

In terms of the temporal logic task, we require that the chaser will arrive a closer place to the target with a lower speed in 60 instants (i.e., 30 minutes), i.e., $P = \{(x,y,v_x,v_y) \mid x \!\in\! [-5,0] \wedge y \!\in\! [-5,0] \wedge v_x \!\in\! [0,2.5] \wedge v_y \!\in\! [0,2.5] \}$, and once in $P$, we can use high precision camera with more state information to monitor, not just position and velocity.
Also, the chaser must always remain with a line-of-sight cone $L = \{(x,y) \mid (y \geq x \tan(30^{\circ})) \wedge (-y \geq x \tan(30^{\circ}))\}$ to keep a good position for docking preparation. 
Such a task can be described by the following formula 
\[
	\Phi = \Phi_1 \wedge \Phi_2,	
\]
where  $\Phi_1 = \mathbf{F}_{[0,60]} (x_k \!\in\! P)$ and $\Phi_2 = \mathbf{G}_{[0,60]} ((x,y) \!\in\! L)$.

Considering a tracjectory of a spacecraft starting from a rectangle points up to $k=19$ instants (i.e., 9.5 minutes) as shown in Fig.~\ref{fig:case4}, the monitor can make sure that the spacecraft will not be able to reach the area that the camera can monitor on time since $x_{19} \!\notin\! X_{19}^{\{1,2\}}$.
This will provide engineers   more time to reschedule the subsequent mission.

\section{Conclusion}\label{sec:con}
In this paper, we proposed a new model-based approach for online monitoring of tasks described by  signal temporal logic formulae  called model predictive monitoring, where we assume the underlying system model is known. 
Our algorithm  consists of both offline pre-computation and online monitoring. 
Most of the computation efforts are made for the pre-computation characterized by the notion of feasible and satisfiable sets. The pre-computation information is used during the online monitoring to provide evaluations in real-time. We showed that the proposed method can evaluate the violation and satisfaction earlier than existing model-free approaches.  Simulation results were provided to illustrate our results.
Note that, in this work, we only consider the STL formula in the form of conjunction of sub-formulae in which the temporal operator will only be applied once for Boolean formulae.
In the future, we would like to consider more general formula, including nested temporal operators, to further generalize our results. Also, we would like to extend our approach to the quantitative setting by taking robustness degrees into account.

\section*{Acknowledgement}
The authors would like to thank Christian Schilling at Aalborg University and Marcelo Forets at Unversidad Tecnológica del Uruguay, two of main contributors of JuliaReach, for their great helps when completing our case studies.

\bibliographystyle{plain}
\bibliography{STL}     

\begin{thebibliography}{10}

\bibitem{abate2019monitor}
M.~Abate, E.~Feron, and S.~Coogan.
\newblock Monitor-based runtime assurance for temporal logic specifications.
\newblock In {\em IEEE Conference on Decision and Control}, pages 1997--2002, 2019.

\bibitem{abbas2022leveraging}
H.~Abbas and B.~Bonakdarpour.
\newblock Leveraging system dynamics in runtime verification of cyber-physical systems.
\newblock In {\em Leveraging Applications of Formal Methods, Verification and Validation. Verification Principles: 11th International Symposium, ISoLA 2022, Rhodes, Greece, October 22--30, 2022, Proceedings, Part I}, pages 264--278. Springer, 2022.

\bibitem{bae2019bounded}
K.~Bae and J.~Lee.
\newblock Bounded model checking of signal temporal logic properties using syntactic separation.
\newblock {\em Proceedings of the ACM on Programming Languages}, 3(POPL):1--30, 2019.

\bibitem{bartocci2018specification}
E.~Bartocci, J.~Deshmukh, A.~Donz{\'e}, G.~Fainekos, O.~Maler, D.~Ni{\v{c}}kovi{\'c}, and S.~Sankaranarayanan.
\newblock Specification-based monitoring of cyber-physical systems: a survey on theory, tools and applications.
\newblock In {\em Lectures on Runtime Verification}, pages 135--175. 2018.

\bibitem{bauer2011runtime}
A.~Bauer, M.~Leucker, and C.~Schallhart.
\newblock Runtime verification for {LTL} and {TLTL}.
\newblock {\em ACM Transactions on Software Engineering and Methodology}, 20(4):1--64, 2011.

\bibitem{benet2019taylormodels}
L.~Benet, M.~Forets, D.~Sanders, and C.~Schilling.
\newblock Taylor{M}odels. jl: Taylor models in {J}ulia and their application to validated solutions of {ODE}s.
\newblock In {\em SWIM}. 2019.

\bibitem{bezanson2017julia}
J.~Bezanson, A.~Edelman, S.~Karpinski, and V.~Shah.
\newblock Julia: A fresh approach to numerical computing.
\newblock {\em SIAM Review}, 59(1):65--98, 2017.

\bibitem{blanchini1994ultimate}
F.~Blanchini.
\newblock Ultimate boundedness control for uncertain discrete-time systems via set-induced {Lyapunov} functions.
\newblock {\em IEEE Transactions on Automatic Control}, 39(2):428--433, 1994.

\bibitem{bogomolov2019juliareach}
S.~Bogomolov, M.~Forets, G.~Frehse, K.~Potomkin, and C.~Schilling.
\newblock Julia{R}each: a toolbox for set-based reachability.
\newblock In {\em ACM International Conference on Hybrid Systems: Computation and Control}, pages 39--44, 2019.

\bibitem{bonnah2022runtime}
E.~Bonnah and K.~Hoque.
\newblock Runtime monitoring of time window temporal logic.
\newblock {\em IEEE Robotics and Automation Letters}, 7(3):5888--5895, 2022.

\bibitem{bravo2005computation}
J.~Bravo, D.~Lim{\'o}n, T.~Alamo, and E.~Camacho.
\newblock On the computation of invariant sets for constrained nonlinear systems: An interval arithmetic approach.
\newblock {\em Automatica}, 41(9):1583--1589, 2005.

\bibitem{buyukkocak2022control}
A.~Buyukkocak, D.~Aksaray, and Y.~Yaz{\i}c{\i}o{\u{g}}lu.
\newblock Control barrier functions with actuation constraints under signal temporal logic specifications.
\newblock In {\em European Control Conference}, pages 162--168. IEEE, 2022.

\bibitem{chan2017verifying}
N.~Chan and S.~Mitra.
\newblock Verifying safety of an autonomous spacecraft rendezvous mission.
\newblock {\em EPiC Series in Computing}, 48:20--32, 2017.

\bibitem{deshmukh2017robust}
J.~Deshmukh, A.~Donz{\'e}, S.~Ghosh, X.~Jin, G.~Juniwal, and S.~Seshia.
\newblock Robust online monitoring of signal temporal logic.
\newblock {\em Formal Methods in System Design}, 51(1):5--30, 2017.

\bibitem{devonport2020data}
A.~Devonport and M.~Arcak.
\newblock Data-driven reachable set computation using adaptive gaussian process classification and monte carlo methods.
\newblock In {\em American Control Conference}, pages 2629--2634. IEEE, 2020.

\bibitem{dokhanchi2014line}
A.~Dokhanchi, B.~Hoxha, and G.~Fainekos.
\newblock On-line monitoring for temporal logic robustness.
\newblock In {\em International Conference on Runtime Verification}, pages 231--246. Springer, 2014.

\bibitem{donze2013efficient}
A.~Donz{\'e}, T.~Ferrere, and O.~Maler.
\newblock Efficient robust monitoring for {STL}.
\newblock In {\em International Conference on Computer Aided Verification}, pages 264--279, 2013.

\bibitem{donze2010robust}
A.~Donz{\'e} and O.~Maler.
\newblock Robust satisfaction of temporal logic over real-valued signals.
\newblock In {\em International Conference on Formal Modeling and Analysis of Timed Systems}, pages 92--106. Springer, 2010.

\bibitem{eisner2003reasoning}
C.~Eisner, D.~Fisman, J.~Havlicek, Y.~Lustig, A.~McIsaac, and D.~Campenhout.
\newblock Reasoning with temporal logic on truncated paths.
\newblock In {\em International Conference on Computer Aided Verification}, pages 27--39, 2003.

\bibitem{evtushenko2018approximating}
Y.~Evtushenko, M.~Posypkin, L.~Rybak, and A.~Turkin.
\newblock Approximating a solution set of nonlinear inequalities.
\newblock {\em Journal of Global Optimization}, 71:129--145, 2018.

\bibitem{fainekos2009robustness}
G.~Fainekos and G.~Pappas.
\newblock Robustness of temporal logic specifications for continuous-time signals.
\newblock {\em Theoretical Computer Science}, 410(42):4262--4291, 2009.

\bibitem{ferrando2022bridging}
A.~Ferrando, R.~Cardoso, M.~Farrell, M.~Luckcuck, F.~Papacchini, M.~Fisher, and V.~Mascardi.
\newblock Bridging the gap between single-and multi-model predictive runtime verification.
\newblock {\em Formal Methods in System Design}, pages 1--33, 2022.

\bibitem{lazysets21}
M.~Forets and C.~Schilling.
\newblock {LazySets.jl: Scalable Symbolic-Numeric Set Computations}.
\newblock {\em Proceedings of the JuliaCon Conferences}, 1(1):11, 2021.

\bibitem{ghosh2022offline}
B.~Ghosh and {\'E}.~Andr{\'e}.
\newblock Offline and online monitoring of scattered uncertain logs using uncertain linear dynamical systems.
\newblock {\em arXiv preprint arXiv:2204.11505}, 2022.

\bibitem{gilpin2020smooth}
Y.~Gilpin, V.~Kurtz, and H.~Lin.
\newblock A smooth robustness measure of signal temporal logic for symbolic control.
\newblock {\em IEEE Control Systems Letters}, 5(1):241--246, 2020.

\bibitem{hashimoto2022stl2vec}
W.~Hashimoto, K.~Hashimoto, and S.~Takai.
\newblock {STL}2vec: Signal temporal logic embeddings for control synthesis with recurrent neural networks.
\newblock {\em IEEE Robotics and Automation Letters}, 2022.

\bibitem{ho2014online}
H.~Ho, J.~Ouaknine, and J.~Worrell.
\newblock Online monitoring of metric temporal logic.
\newblock In {\em International Conference on Runtime Verification}, pages 178--192, 2014.

\bibitem{jagtap2020formal}
P.~Jagtap, S.~Soudjani, and M.~Zamani.
\newblock Formal synthesis of stochastic systems via control barrier certificates.
\newblock {\em IEEE Transactions on Automatic Control}, 66(7):3097--3110, 2020.

\bibitem{jakvsic2018quantitative}
S.~Jak{\v{s}}i{\'c}, E.~Bartocci, R.~Grosu, T.~Nguyen, and D.~Ni{\v{c}}kovi{\'c}.
\newblock Quantitative monitoring of {STL} with edit distance.
\newblock {\em Formal Methods in System Design}, 53(1):83--112, 2018.

\bibitem{kerrigan2001robust}
E.~Kerrigan.
\newblock {\em Robust constraint satisfaction: Invariant sets and predictive control}.
\newblock PhD thesis, University of Cambridge, 2001.

\bibitem{lee2021efficient}
J.~Lee, G.~Yu, and K.~Bae.
\newblock Efficient smt-based model checking for signal temporal logic.
\newblock In {\em 2021 36th IEEE/ACM International Conference on Automated Software Engineering (ASE)}, pages 343--354. IEEE, 2021.

\bibitem{leucker2012sliding}
M.~Leucker.
\newblock Sliding between model checking and runtime verification.
\newblock In {\em International Conference on Runtime Verification}, pages 82--87. Springer, 2012.

\bibitem{lindemann2017robust}
L.~Lindemann and D.~Dimarogonas.
\newblock Robust motion planning employing signal temporal logic.
\newblock In {\em American Control Conference}, pages 2950--2955. IEEE, 2017.

\bibitem{lindemann2018control}
L.~Lindemann and D.~Dimarogonas.
\newblock Control barrier functions for signal temporal logic tasks.
\newblock {\em IEEE Control Systems Letters}, 3(1):96--101, 2018.

\bibitem{lindemann2019robust}
L.~Lindemann and D.~Dimarogonas.
\newblock Robust control for signal temporal logic specifications using discrete average space robustness.
\newblock {\em Automatica}, 101:377--387, 2019.

\bibitem{lindemann2022conformal}
L.~Lindemann, X.~Qin, J.~Deshmukh, and G.~Pappas.
\newblock Conformal prediction for {STL} runtime verification.
\newblock {\em arXiv preprint arXiv:2211.01539}, 2022.

\bibitem{ma2021novel}
M.~Ma, E.~Bartocci, E.~Lifland, J.~Stankovic, and L.~Feng.
\newblock A novel spatial--temporal specification-based monitoring system for smart cities.
\newblock {\em IEEE Internet of Things Journal}, 8(15):11793--11806, 2021.

\bibitem{ma2021predictive}
M.~Ma, J.~Stankovic, E.~Bartocci, and L.~Feng.
\newblock Predictive monitoring with logic-calibrated uncertainty for cyber-physical systems.
\newblock {\em ACM Transactions on Embedded Computing Systems}, 20(5s):1--25, 2021.

\bibitem{maler2004monitoring}
O.~Maler and D.~Nickovic.
\newblock Monitoring temporal properties of continuous signals.
\newblock In {\em Formal Techniques, Modelling and Analysis of Timed and Fault-Tolerant Systems}, pages 152--166. 2004.

\bibitem{mascle2020ltl}
C.~Mascle, D.~Neider, M.~Schwenger, P.~Tabuada, A.~Weinert, and M.~Zimmermann.
\newblock From {LTL} to {rLTL} monitoring: Improved monitorability through robust semantics.
\newblock In {\em International Conference on Hybrid Systems: Computation and Control}, pages 1--12, 2020.

\bibitem{mayne2001control}
D.~Mayne.
\newblock Control of constrained dynamic systems.
\newblock {\em European Journal of Control}, 7(2-3):87--99, 2001.

\bibitem{mitchell2003overapproximating}
I.~Mitchell and C.~Tomlin.
\newblock Overapproximating reachable sets by hamilton-jacobi projections.
\newblock {\em journal of Scientific Computing}, 19:323--346, 2003.

\bibitem{momtaz2021predicate}
A.~Momtaz, N.~Basnet, H.~Abbas, and B.~Bonakdarpour.
\newblock Predicate monitoring in distributed cyber-physical systems.
\newblock In {\em International Conference on Runtime Verification}, pages 3--22. Springer, 2021.

\bibitem{pinisetty2017predictive}
S.~Pinisetty, T.~J{\'e}ron, S.~Tripakis, Y.~Falcone, H.~Marchand, and V.~Preoteasa.
\newblock Predictive runtime verification of timed properties.
\newblock {\em Journal of Systems and Software}, 132:353--365, 2017.

\bibitem{qin2020clairvoyant}
X.~Qin and J.~Deshmukh.
\newblock Clairvoyant monitoring for signal temporal logic.
\newblock In {\em International Conference on Formal Modeling and Analysis of Timed Systems}, pages 178--195. Springer, 2020.

\bibitem{raman2014model}
V.~Raman, A.~Donz{\'e}, M.~Maasoumy, R.~Murray, A.~Sangiovanni-Vincentelli, and S.~Seshia.
\newblock Model predictive control with signal temporal logic specifications.
\newblock In {\em IEEE Conference on Decision and Control}, pages 81--87, 2014.

\bibitem{roehm2016stl}
H.~Roehm, J.~Oehlerking, T.~Heinz, and M.~Althoff.
\newblock {STL} model checking of continuous and hybrid systems.
\newblock In {\em Automated Technology for Verification and Analysis: 14th International Symposium, ATVA 2016, Chiba, Japan, October 17-20, 2016, Proceedings 14}, pages 412--427. Springer, 2016.

\bibitem{sahin2020autonomous}
Y.~Sahin, R.~Quirynen, and S.~Di~Cairano.
\newblock Autonomous vehicle decision-making and monitoring based on signal temporal logic and mixed-integer programming.
\newblock In {\em American Control Conference}, pages 454--459. IEEE, 2020.

\bibitem{salamati2021data}
A.~Salamati, S.~Soudjani, and M.~Zamani.
\newblock Data-driven verification of stochastic linear systems with signal temporal logic constraints.
\newblock {\em Automatica}, 131:109781, 2021.

\bibitem{stipanovic2003computation}
D.~Stipanovi{\'c}, I.~Hwang, and C.~Tomlin.
\newblock Computation of an over-approximation of the backward reachable set using subsystem level set functions.
\newblock In {\em 2003 European Control Conference (ECC)}, pages 300--305. IEEE, 2003.

\bibitem{thati2005monitoring}
P.~Thati and G.~Ro{\c{s}}u.
\newblock Monitoring algorithms for metric temporal logic specifications.
\newblock {\em Electronic Notes in Theoretical Computer Science}, 113:145--162, 2005.

\bibitem{waga2021model}
M.~Waga, {\'E}.~Andr{\'e}, and I.~Hasuo.
\newblock Model-bounded monitoring of hybrid systems.
\newblock In {\em ACM/IEEE International Conference on Cyber-Physical Systems}, pages 21--32, 2021.

\bibitem{yoon2021predictive}
H.~Yoon and S.~Sankaranarayanan.
\newblock Predictive runtime monitoring for mobile robots using logic-based {B}ayesian intent inference.
\newblock In {\em IEEE International Conference on Robotics and Automation}, pages 8565--8571. IEEE, 2021.

\bibitem{yu2022stlmc}
G.~Yu, J.~Lee, and K.~Bae.
\newblock {STL}mc: Robust {STL} model checking of hybrid systems using {SMT}.
\newblock In {\em Computer Aided Verification: 34th International Conference, CAV 2022, Haifa, Israel, August 7--10, 2022, Proceedings, Part I}, pages 524--537. Springer, 2022.

\bibitem{yu2022online}
X.~Yu, W.~Dong, X.~Yin, and S.~Li.
\newblock Online monitoring of dynamic systems for signal temporal logic specifications with model information.
\newblock In {\em IEEE Conference on Decision and Control}, pages 1553--1559. IEEE, 2022.

\bibitem{xiao2022decentralized}
X.~Yu and R.~Su.
\newblock Decentralized circular formation control of nonholonomic mobile robots under a directed sensor graph.
\newblock {\em IEEE Transactions on Automatic Control}, pages 1--8, 2022.

\bibitem{zhang2012runtime}
X.~Zhang, M.~Leucker, and W.~Dong.
\newblock Runtime verification with predictive semantics.
\newblock In {\em NASA Formal Methods Symposium}, pages 418--432. Springer, 2012.

\bibitem{zhao2022astl}
D.~Zhao, Z.~Zhou, Z.~Cai, S.~Yangui, and X.~Xue.
\newblock {ASTL}: Accumulative {STL} with a novel robustness metric for iot service monitoring.
\newblock {\em IEEE Transactions on Mobile Computing}, 2022.

\end{thebibliography}

\end{document}